\documentclass[osajnl,twocolumn,showpacs,superscriptaddress,10pt]{revtex4-1} 
\usepackage{amsmath,amssymb,graphicx}

\begin{document}

\title{Non-diffracting and non-attenuating vortex light beams in media \\with nonlinear absorption of orbital angular momentum}

\author{Miguel A. Porras}\email{Corresponding author: miguelangel.porras@upm.es}
\affiliation{Departamento de F\'{i}sica Aplicada a los Recursos Naturales and Grupo de Sistemas Complejos,
Universidad Polit\'{e}cnica de Madrid, Rios Rosas 21, Madrid ES-28003, Spain}

\author{Carlos Ruiz-Jim\'enez}
\affiliation{Grupo Sistemas Complejos, Universidad Polit\'ecnica de Madrid, 28040 Madrid, Spain}

\begin{abstract}
We show that a high-order Bessel beam propagating in a medium with nonlinear absorption is not completely absorbed, but survives in the form of a new propagation invariant vortex beam in which the beam energy and orbital angular momentum are permanently transferred to matter and at the same time refueled by spiral inward currents of energy and angular momentum. Unlike vortex solitons and dissipative vortex solitons, these vortex beams are not supported by specific dispersive nonlinearities (self-focusing or self-defocusing) and do not require gain. Propagation invariance in presence of multiple absorption of photons carrying (possibly high) orbital angular momentum makes these beams attractive for optical pumping of angular momentum over long distances.
\end{abstract}


\maketitle

\section{Introduction}

There is a sustained interest in beams of light carrying orbital angular momentum, particularly in those with screw topological
wave front dislocations or vortices. These vortex light beams have open new perspectives in information encoding, quantum entanglement, imaging, optical trapping, and in diverse forms of transference of optical angular momentum to matter, e. g., to micro- or nano-particles, to Bose-Einstein condensates or atoms \cite{ALLEN,TABOSA, MALASOV}. Although most of the research initially focused on Laguerre-Gauss beams \cite{LG}, many other vortex beams have been explored, many of them endowed of the advantageous property of being diffraction-free, as high-order Bessel beams (BBs) in linear media \cite{BB,BB2}, vortex solitons in transparent media with suitable nonlinearities \cite{SL,SELFDEFOCUSING,CQ,NONLOCAL}, or dissipative vortex solitons in nonlinear media with gain and losses \cite{DSL,DSL2,DSL3}.

In applications involving pumping of orbital angular momentum to matter, the net transference of energy and angular momentum implies a power loss of the pumping vortex beam, whose effect in its propagation is not usually taken into account. In (dissipative) vortex solitons, for example, perturbation of the precise balance between nonlinearities and diffraction and between gain and losses may result in the quenching of the solitary regime.

Although there is no known light beam that can overcome the attenuation effects of linear absorption, there are light beams, known as nonlinear unbalanced Bessel beams (NL-UBBs) that can propagate without any diffraction and attenuation while their energy is transferred to matter nonlinearly via multiphoton absorption \cite{PORRAS1}. NL-UBBs belong to the family of conical waves and as such they transport infinite power. This power reservoir flows permanently from the linear outer rings towards the central peak of intensity, where most of nonlinear power losses take place, refilling it. NL-UBBs has been shown to play an important role in the filamentation of ultrahsort pulses, \cite{PORRAS1,SF} to be able to create long-lived fluorescence channels excited by multiphoton absorption of the NL-UBB power \cite{CHANNELS}, and to be easily generated experimentaly from a zero-order linear BB \cite{CONE}.

In this paper, we point out that the above NL-UBB is the fundamental member of an infinite family of NL-UBBs with helical wave fronts and carrying orbital angular momentum. Unlike linear vortex beams and vortex solitons, these vortex NL-UBBs incorporate a mechanism of transference of their angular momentum to matter: the nonlinear absorption just supporting their stationarity. Light power and angular momentum spiral inward permanently, and they are transferred to matter within a thin ring surrounding the phase singularity. Simultaneous absorption of $M$ photons, each photon carrying (arbitrarily high) orbital angular momentum $\hbar l$ ($l=\pm 1, \pm 2,\dots $) could be used as an efficient method of optical pumping of angular momentum \cite{MALASOV,TABOSA}. We also stress that self-focusing or self-defocusing nonlinearities \cite{SL,SELFDEFOCUSING}, saturable or not \cite{CQ}, local or non-local \cite{NONLOCAL}, are accessory nonlinearities for vortex NL-UBBs that do not contribute substantially to the their propagation invariance property, and that unlike dissipative vortex solitons, a balancing gain is not required.

Experiments in Ref. \cite{CONE} demonstrated that a zero-order BB launched in a nonlinear medium reshapes spontaneously into a fundamental NL-UBB of the same cone angle as the input BB and of certain peak intensity. Here we show that a high-order BB transforms similarly into a vortex NL-UBB that preserves the cone angle and the topological charge. We further find a third preserved quantity of the propagation that allows us to provide with a solution to the so-called ``selection problem," or determination the specific (vortex or vortex-less) NL-UBB that is selected as the final stage of the nonlinear propagation of the input BB, a problem that remained open in Ref. \cite{CONE}.

\section{Propagation of high-order Bessel beams in nonlinear media with nonlinear losses}\label{DYNAMICS1}

Accordingly, we consider a light beam $E=A\exp[-i(\omega t-kz)]$ of angular frequency $\omega$ and linear propagation constant $k=(\omega/c)n$, where $n$ is the linear refractive index and $c$ is the speed of light in vacuum. In the paraxial approximation the change of the complex envelope $A$ along the propagation direction $z$ can be described by a nonlinear Schr\"odinger equation (NLSE) of the type
\begin{equation}\label{NLSE}
\partial_z A =\frac{i}{2k}\Delta_\perp A + if(|A|^2)A - \frac{\beta^{(M)}}{2}|A|^{2M-2}A \,,
\end{equation}
where $\Delta_\perp=\partial^2_r+(1/r)\partial_r+(1/r^2)\partial^2_\varphi$ is the transversal Laplace operator, $(r,\varphi)$ are polar coordinates in the transversal plane, $\beta^{(M)}>0$ is the $M$-photon absorption coefficient, and dispersive nonlinearities as self-focusing or self-defocusing are included in the term with $f(|A|^2)$.

In absence of all nonlinear terms, Eq. (\ref{NLSE}) is satisfied by the high-order (vortex) BB $A(r,\varphi,z) \propto J_l(\sqrt{2k|\delta|} r)e^{il\varphi}e^{i\delta z}$ with any $\delta <0$, of cone angle $\theta= \sqrt{2|\delta|/k}$ and topological charge $l=\pm 1, \pm 2 \dots$. For a more comprehensive analysis of the propagation of BBs in nonlinear media, we introduce dimensionless radius $\rho=\sqrt{2k|\delta|} r$, propagation distance $\zeta=|\delta|z$ and envelope $\tilde A=(\beta^{(M)}/2|\delta|)^{1/(2M-2)}A$, with which Eq. (\ref{NLSE}) rewrites as
\begin{equation}\label{NLSEN}
\partial_\zeta \tilde A= i\Delta_\perp\tilde A + i \tilde f(|\tilde A|^2)\tilde A - |\tilde A|^{2M-2}\tilde A
\end{equation}
where now $\Delta_\perp=\partial^2_\rho+(1/\rho)\partial_\rho+(1/\rho^2) \partial^2_\varphi$, and the vortex BB launched at the entrance plane $\zeta=0$ of the medium rewrites as $\tilde A(\rho,\varphi,0)= b_0 J_l(\rho)e^{il\varphi}$, where the parameter $b_0$ controls its intensity. For illustration purposes, we consider cubic and quintic dispersive nonlinearities, in which case $f(u)=k(n_2 u + n_4 u^2)/n$ in Eq. (\ref{NLSE}), where  $n_j$ ($j=2,4$) are the nonlinear refractive indexes, and $\tilde f(u)=\alpha_2 u + \alpha_4 u^2$ in Eq. (\ref{NLSEN}), where $\alpha_j= (2|\delta|/\beta^{(M)})^{j/(2M-2)}kn_j/(n|\delta|)$.
\begin{center}
\begin{figure}
\includegraphics[width=6.2cm]{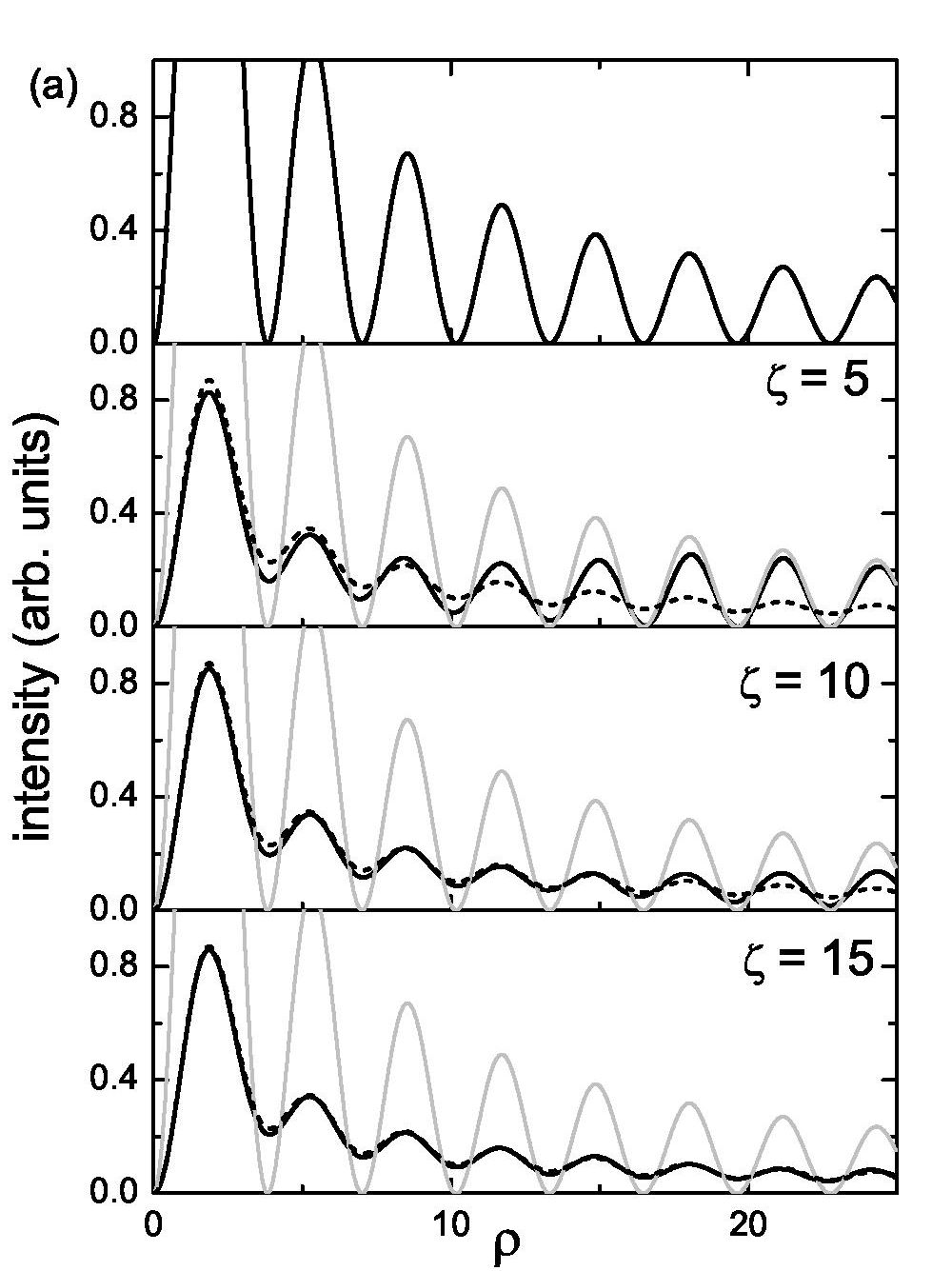}
\includegraphics[width=6.2cm]{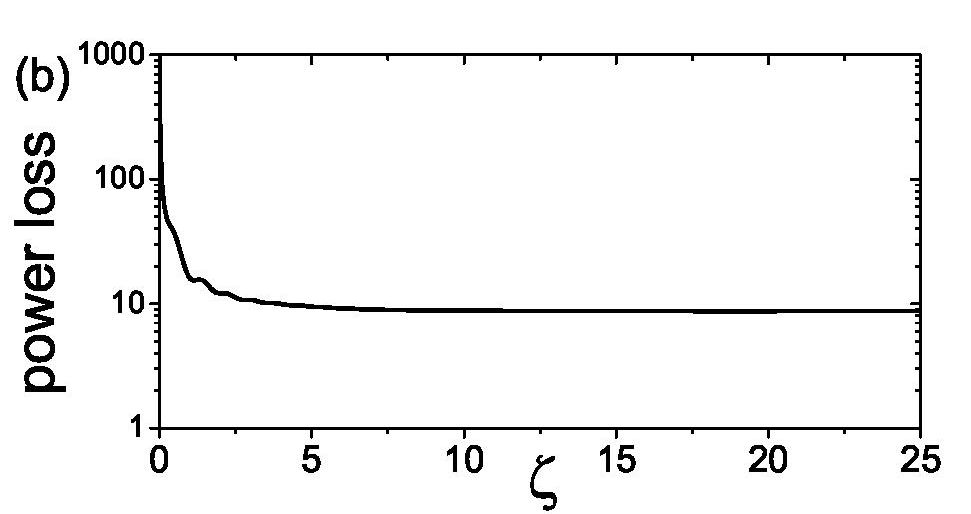}
\caption{\label{Fig1} (a) Nonlinear dynamics of an vortex BB $\tilde A=b_0 J_l\rho)e^{il\varphi}$ with topological charge $l=1$, amplitude $b_0=3$ in a medium with four-photon absorption $M=4$ and $\alpha_2 = 0$ and $\alpha_4 = 0$.
The gray curves represent the intensity of the input vortex BB, and the dashed curves the intensity of the propagation invariant beam reached at large $\zeta$ (the vortex NL-UBB with $b_0=1.60$, $l=1$ and $M=4$). (b) Nonlinear power losses per unit propagation length as a function of propagation distance.}
\end{figure}
\end{center}
\begin{center}
\begin{figure}
\includegraphics[width=6.2cm]{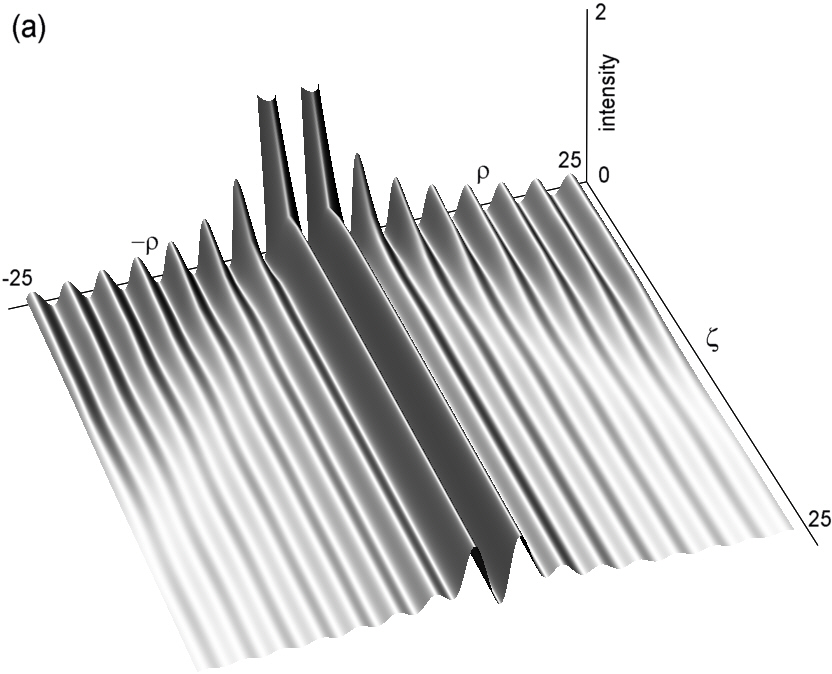}
\includegraphics[width=6.2cm]{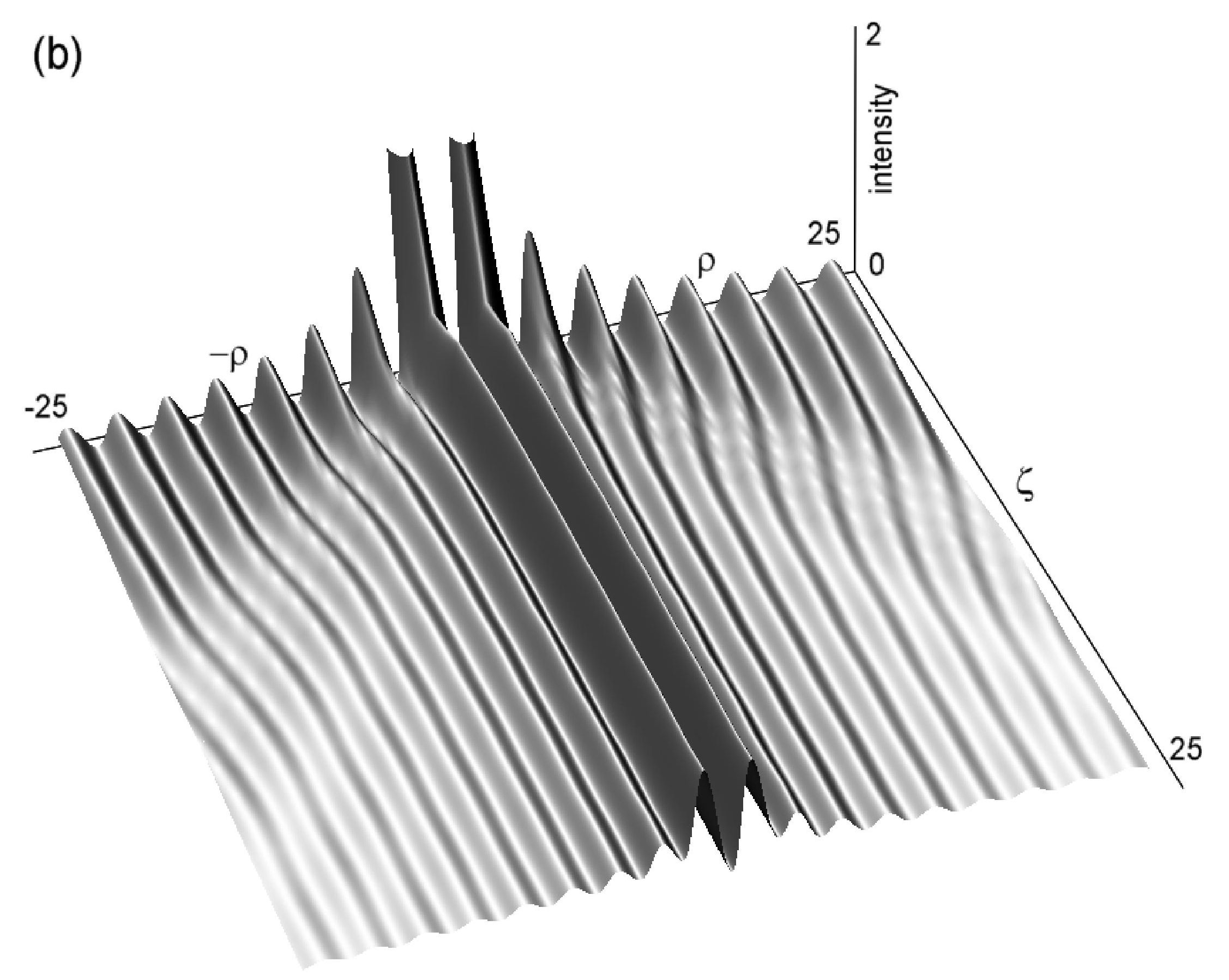}
\includegraphics[width=6.2cm]{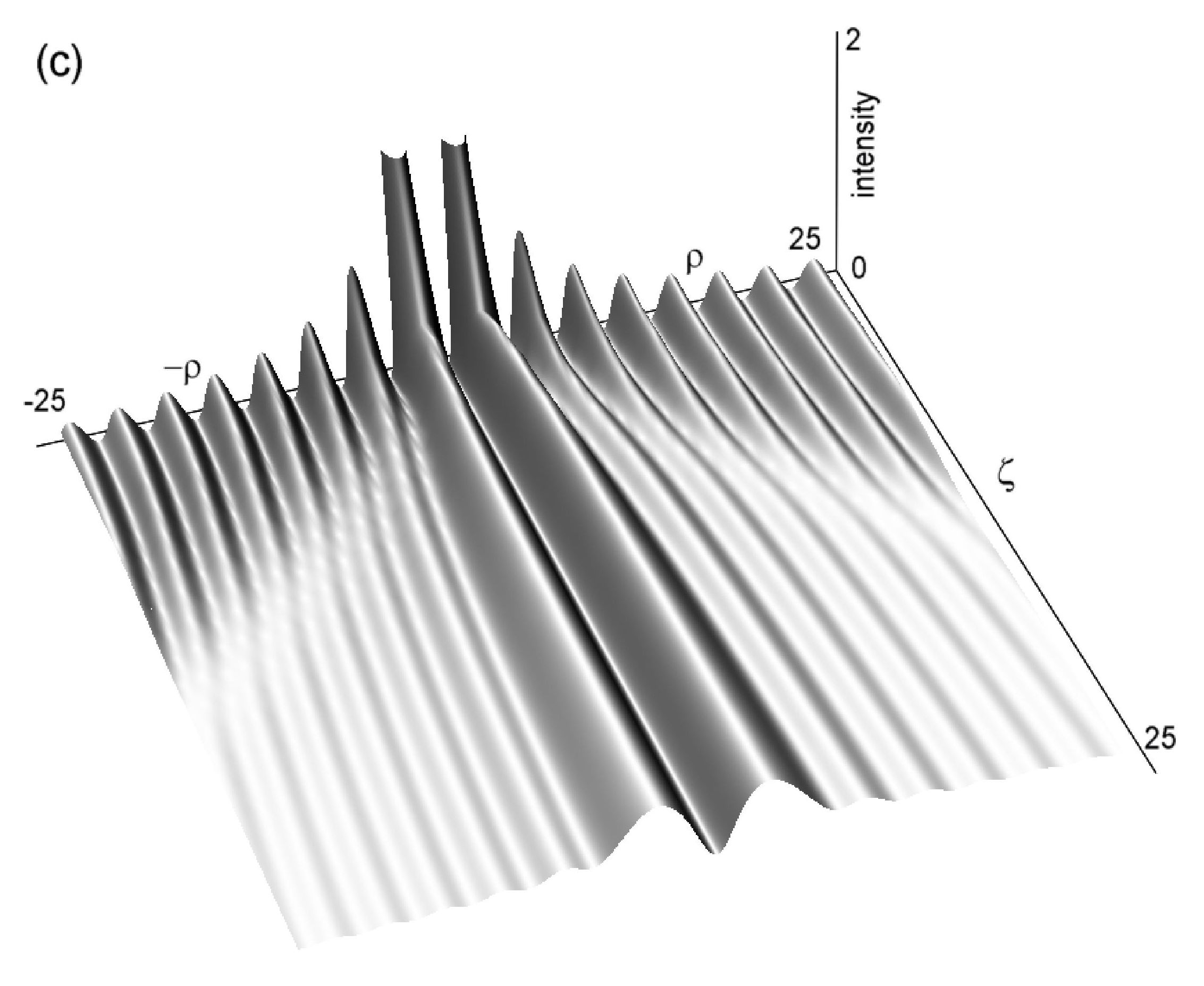}
\caption{\label{Fig2} Nonlinear dynamics of the vortex BB $\tilde A=b_0 J_l(\rho)e^{il\varphi}$ with topological charge $l=1$,  amplitude $b_0=3$ in media with four photon absorption $M=4$ and (a) $\alpha_2 = 0$ and $\alpha_4 = 0$, (b) $\alpha_2=2$ and $\alpha_4=-1$, and (c) $\alpha_2=-2$ and $\alpha_4=1$.}
\end{figure}
\end{center}
Numerical solution of the NLSE in Eq. (\ref{NLSEN}) shows that high-order BBs are not completely depleted by nonlinear absorption, as would happen with a plane wave or with a Gaussian beam, but they transform into new non-diffracting and non-attenuating beams. Figure \ref{Fig1}(a) illustrates the transformation dynamics in the simplest situation of negligible dispersive nonlinearities ($\alpha_j=0$). After an initial stage of sudden absorption (provided $b_0$ is large), the beam profile stabilizes into the new propagation-invariant beam, characterized by the same cone angle and topological charge as the input vortex BB, but with a reduced contrast of the annuli surrounding the vortex. Remarkably, the beam do not attenuate in spite that the nonlinear power losses per unit propagation length
\begin{equation}
N(\infty)=2\pi\int_0^{\infty}d\rho\rho |\tilde A|^{2M}
\end{equation}
in the final propagation-invariant regime beam are constant and non-negligible, as seen in Fig. \ref{Fig1}(b). The radial profile, particularly the contrast of the annuli of the final propagation invariant vortex beam, and also its nonlinear power losses, strongly depend on the amplitude $b_0$ of the input BB. The gradual transformation into the new vortex beam always starts at the beam center, and fills the cone $\rho=2\zeta$ at a propagation distance $\zeta$ (the cone $r=\theta z$ in real-world variables), as can be appreciated in Fig. \ref{Fig2}(a). Similar dynamics is also observed when dispersive nonlinearities are included, e. g., in Fig. \ref{Fig2}(b) for a saturable self-focusing nonlinearity, or in Fig. \ref{Fig2}(c) for a saturable self-defocusing nonlinearity. The only significant difference is that the inner rings in the final stationary regime may be thinner or thicker, and that the outer rings are shifted radially to match the inner ones.

An analogous dynamics has been experimentally observed \cite{CONE} for the transformation of a zero-order Bessel beam into a vortex-less NL-UBB. The above simulations generalize those observations to vortex beams, and evidence the existence of non-diffracting beams that can transfer their energy and orbital angular momentum without any attenuation. According to these simulations, these non-diffracting beams do not require specific dispersive nonlinearities balancing (more or less stably) diffraction spreading, as for vortex solitons in media with competing self-focusing and self-defocusing nonlinearities \cite{CQ}, or with non-local self-focusing nonlinearity \cite{NONLOCAL}.

In Sec. \ref{STATIONARY} we investigate more in-depth on the properties of these vortex NL-UBBs. This will allow us to complete in Sec. \ref{SELECTION} the description of the (vortex or vortex-less) BB nonlinear dynamics by fully specifying the particular propagation invariant NL-UBB that is formed for each particular amplitude $b_0$ of the input BB.

\section{Vortex beams supported by nonlinear losses}\label{STATIONARY}

We above analysis suggests to search for solutions to Eq. (\ref{NLSE}) of the form $A(r,\varphi,z)=a(r)e^{i\phi(r)}e^{il\varphi}e^{i\delta z}$, or in dimensionless variables, solutions to Eq. (\ref{NLSEN}) of the form $\tilde A(\rho,\varphi,\zeta)=\tilde a(\rho)e^{i\phi(\rho)}e^{il\varphi}e^{ \pm i\zeta}$, where $\pm$ is the sign of $\delta$. According to Eq. (\ref{NLSEN}) the amplitude $\hat a(\rho)$ and phase $\phi(\rho)$ profiles must satisfy
\begin{eqnarray}
\frac{d^2\tilde a}{d\rho^2}+\frac{1}{\rho}\frac{d\tilde a}{d\rho}-\left(\frac{d\phi}{d\rho}\right)^2 \tilde a +\tilde f(\tilde a^2) \tilde a - (\pm\tilde a) -\frac{l^2}{\rho^2}\tilde a &=&0 \,, \label{AMP}\\
\frac{d^2\phi}{d\rho^2}+ \frac{1}{\rho}\frac{d\phi}{d\rho} + 2\frac{d\phi}{d\rho}\frac{d\tilde a}{d\rho}\frac{1}{\tilde a} + \tilde a^{2M-2}&=&0\,,
\label{PHAS}
\end{eqnarray}
The vortex of charge $l$ at $\rho=0$ requires the amplitude to behave as $\tilde a(\rho)\propto \rho^l$ as $\rho\rightarrow 0$, and the localization condition requires $\tilde a(\rho)\rightarrow 0$ as $\rho\rightarrow \infty$. In absence of nonlinear absorption, and depending on the specific dispersive nonlinearities, the above boundary problem with the positive sign ($\delta>0$) has a discrete spectrum of solutions [$\tilde a \simeq c\rho^l$ as $\rho\rightarrow 0$ with discrete values of $c>0$], or vortex solitons. With the negative sign ($\delta<0$), the above problem has a continuous spectrum of solutions [$\tilde a\simeq c\rho^l$ for any $c>0$], or vortex nonlinear BBs. Since Eq. (\ref{PHAS}) without the absorption term is satisfied by $\phi=\mbox{const.}$, only Eq. (\ref{AMP}) is usually written explicitly \cite{SL}. The introduction of nonlinear absorption washes out the discrete spectrum in the above problem with the positive sign, i. e., vortex solitons cease to exist in presence of nonlinear absorption. The continuous spectrum for the negative sign, however, continues to exist, but limited to a maximum value of the constant $c$, which constitute the continuous set of vortex NL-UBBs.

\begin{figure}[t]
\begin{center}
\includegraphics[width=4.2cm]{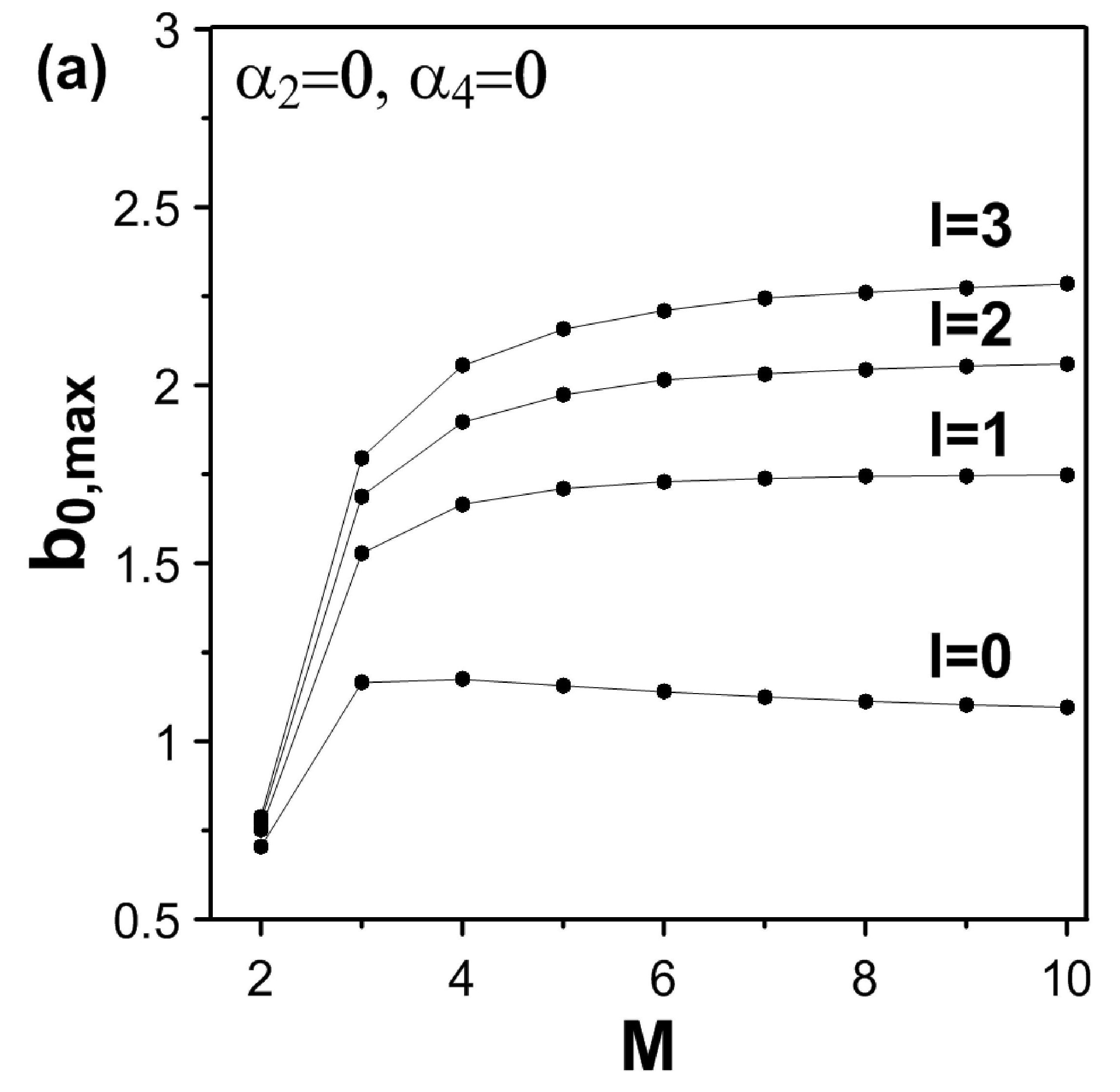}
\includegraphics[width=4.2cm]{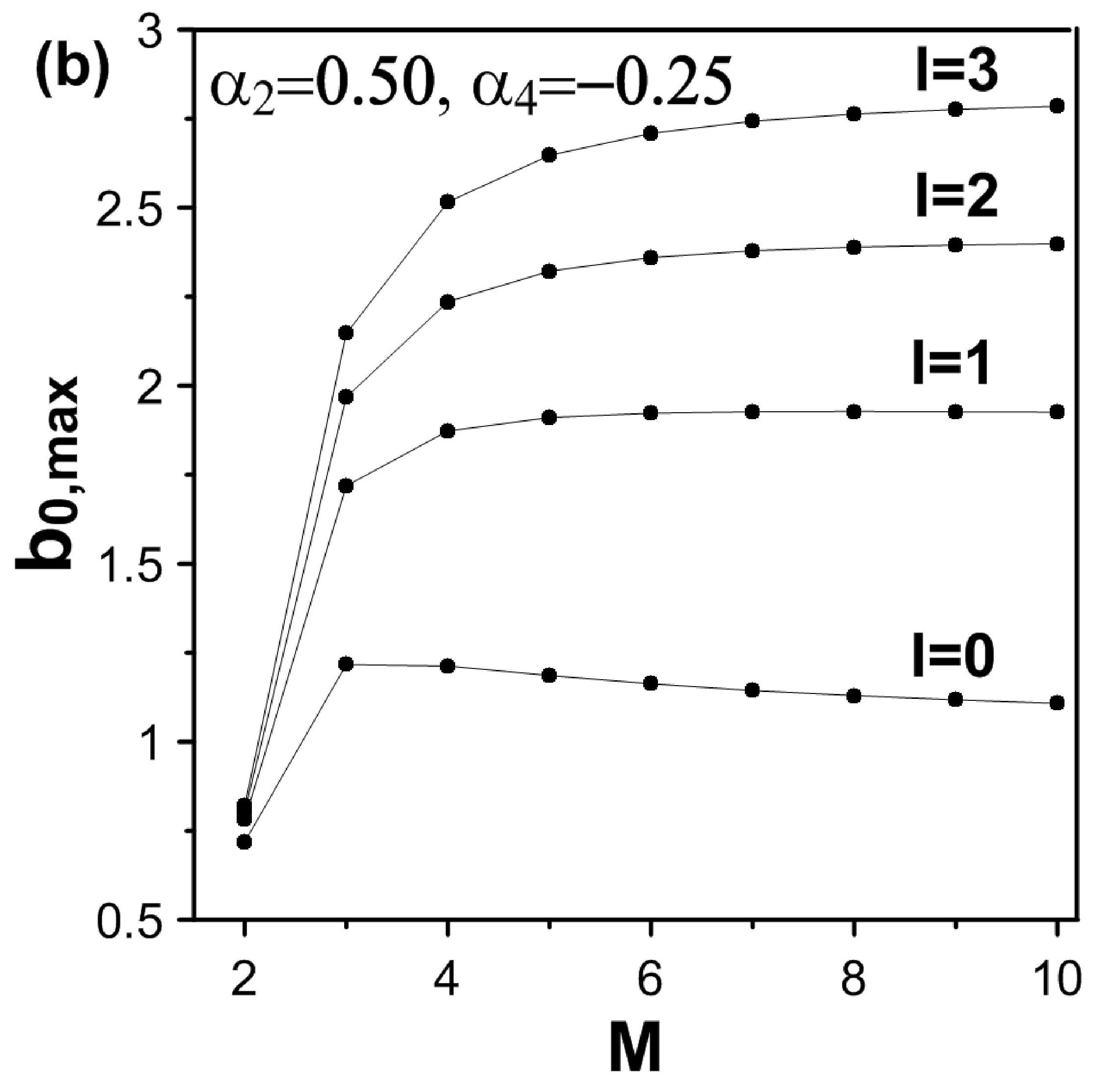}
\end{center}
\caption{\label{Fig3} Values of $b_{0,\rm max}$ of NL-UBBs with different topological charge $l$ in media with different $M$ and with (a) $\alpha_2=\alpha_4=0$, (b) $\alpha_2=0.5$ and $\alpha_4=-0.25$.}
\end{figure}

\begin{figure}[t]
\begin{center}
\includegraphics[width=4.1cm]{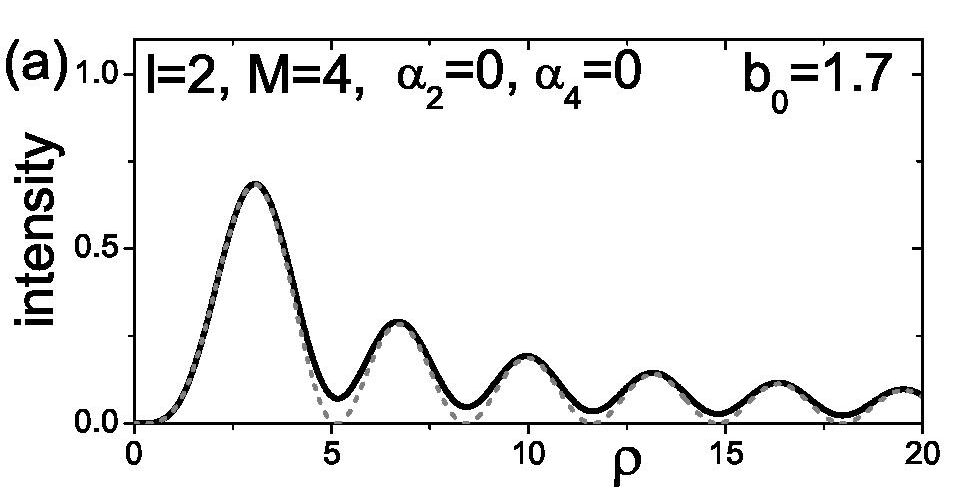}
\includegraphics[width=4.1cm]{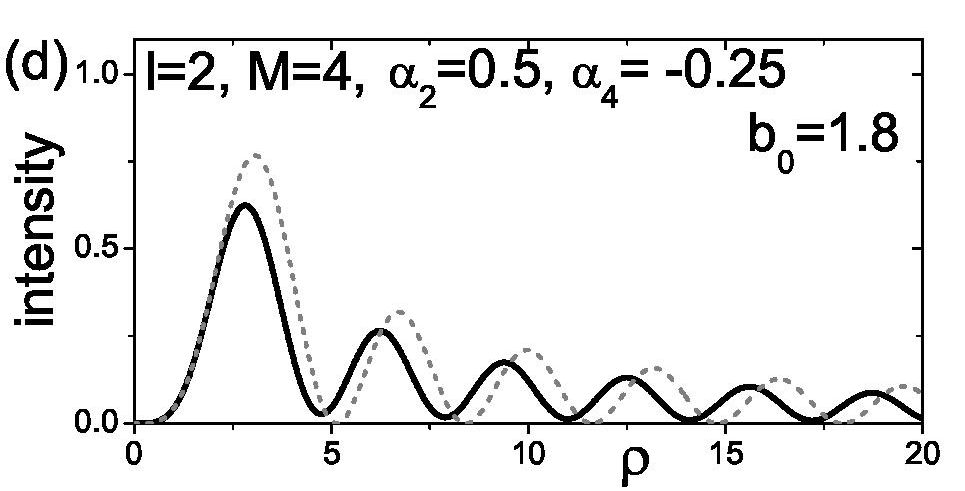}
\includegraphics[width=4.1cm]{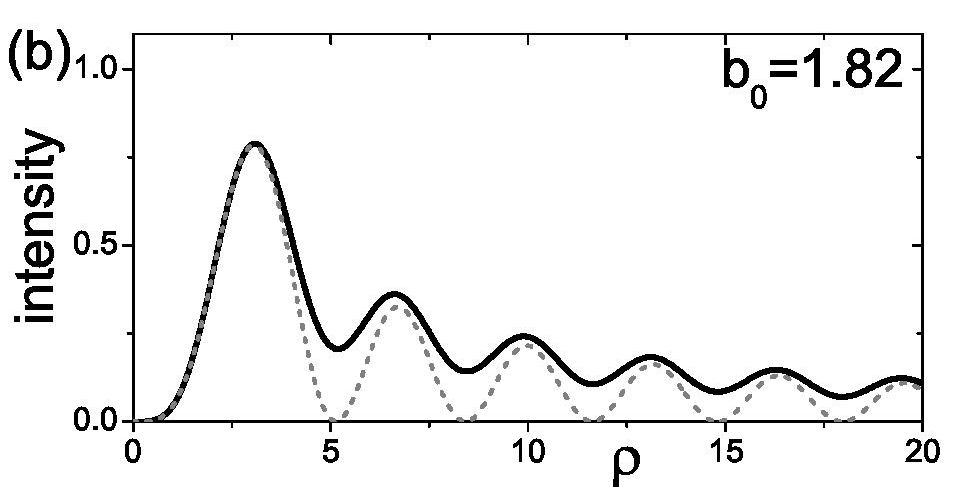}
\includegraphics[width=4.1cm]{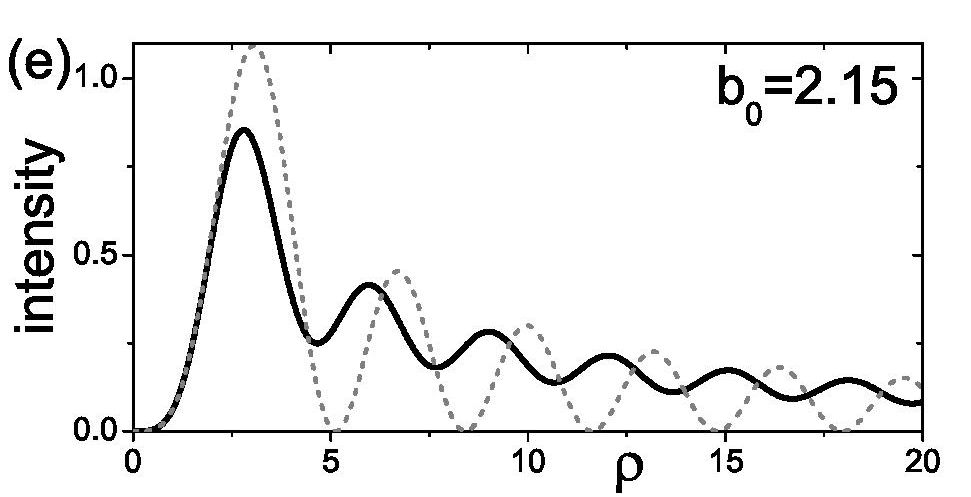}
\includegraphics[width=4.1cm]{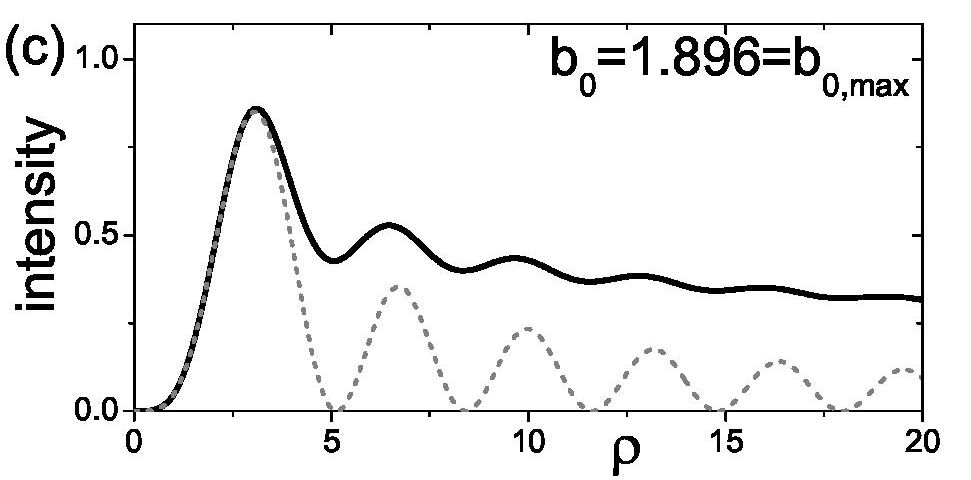}
\includegraphics[width=4.1cm]{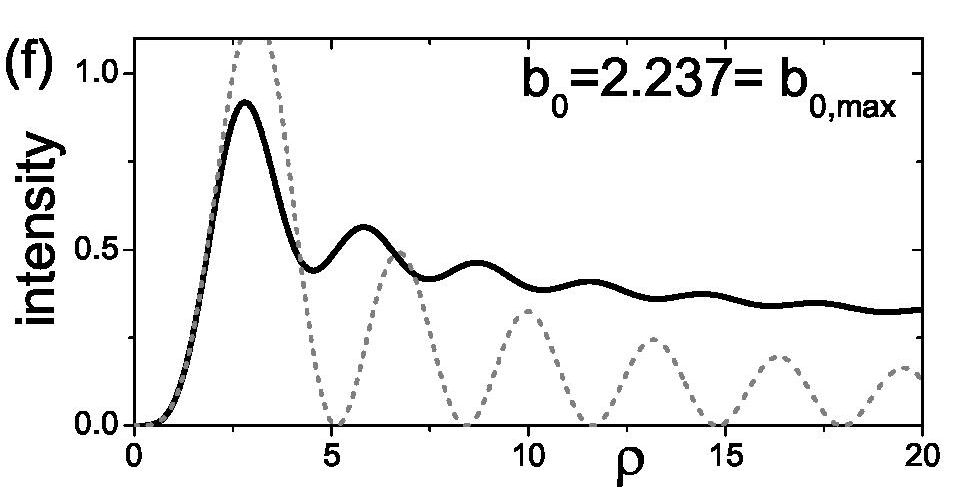}
\end{center}
\caption{\label{Fig4} For $\alpha_2=\alpha_4=0$, intensity profiles of vortex NL-UBBs with $l=2$, $M=4$, and (a) $b_0=1.7$, (b) $b_0=1.82$ and (c) $b_0=1.896$. For $\alpha_2=0.5$, $\alpha_4=-0.25$  intensity profiles of vortex NL-UBBs with $l=2$, $M=4$, and (d) $b_0=1.8$, (e) $b_0=2.15$ and (f) $b_0=2.237$. The dotted curves represent linear vortex BBs with $l=2$ and the same $b_0$ as the vortex NL-UBBs.}
\end{figure}

For a physically meaningful classification of the vortex NLL-UBBs, we first note that all nonlinear terms in Eqs. (\ref{AMP}) and (\ref{PHAS}) are negligible in the vortex core $\rho\rightarrow 0$. Therefore vortex NL-UBBs behave as a linear vortex BB $\tilde a\propto J_l(\rho)$ as $\rho\rightarrow 0$. Since $J_l(\rho)\simeq \rho^l/(2^l l!)$ as $\rho\rightarrow 0$, the solution of the nonlinear problem in Eqs. (\ref{AMP}) and (\ref{PHAS}) satisfying the boundary condition
\begin{eqnarray}\label{B}
\tilde a(\rho)\simeq \frac{b_0}{2^l l!}\rho^l\quad \mbox{as} \quad \rho\rightarrow 0
\end{eqnarray}
represents the vortex NL-UBB whose vortex core is that of the vortex BB $b_0J_l(\rho)$. As said, NL-UBBs in nonlinearly absorbing are seen to exist (satisfy the localization condition) only for $b_0\in(0,b_{0,\rm \max}]$, where $b_{0,\rm max}$ depends on $l$, $M$ and the specific dispersive nonlinearities. Numerically evaluated values of $b_{0,\rm max}$ are displayed in Fig. \ref{Fig3} (a) and (b) as functions of $M$ for different values of $l$, in absence of dispersive nonlinearities and in a particular self-focusing medium. A few examples of their intensity radial profiles $\tilde a^2(\rho)$ are shown in Fig. \ref{Fig4} (solid curves) and compared to the vortex BB with the same vortex core (same $b_0$). At low amplitude (small $b_0$), the NL-UBB behaves as the BB not only at the vortex core but at any radial distance. When $b_0$ increases up to $b_{0,\rm max}$, the contrast of the rings is gradually lost, and the inner rings become compressed (enlarged) in self-focusing (self-defocusing) media. In all cases, the outer rings at $\rho\rightarrow\infty$ decay in amplitude as $\rho^{-1/2}$ and oscillate at the same radial frequency as those of the linear BB. In self-focusing or self-defocusing media, these outer rings are therefore radially shifted with respect to those of the linear vortex BB so that they match the compressed or enlarged inner rings.
\begin{figure*}
\begin{center}
\includegraphics[width=5cm]{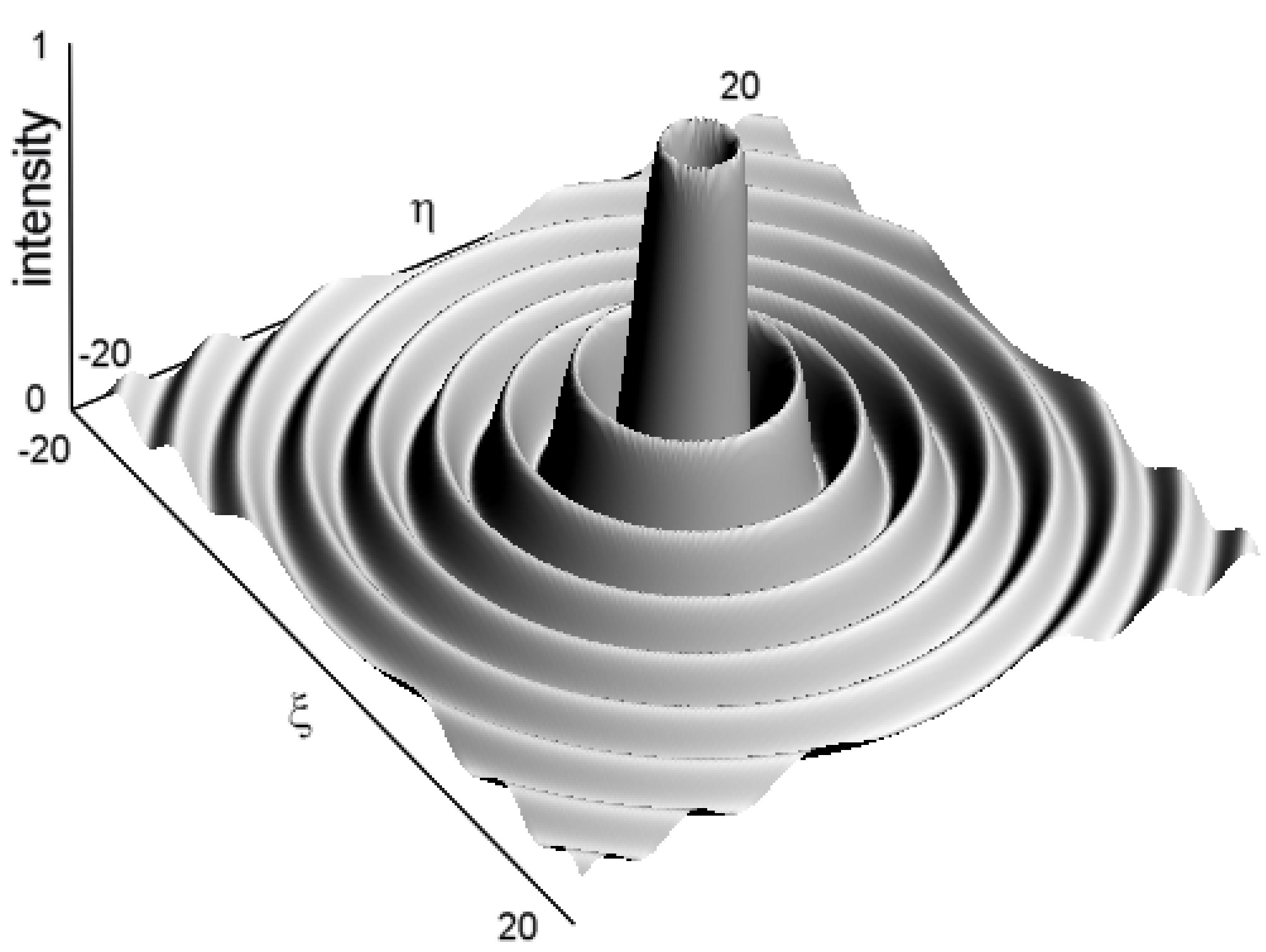}
\includegraphics[width=5cm]{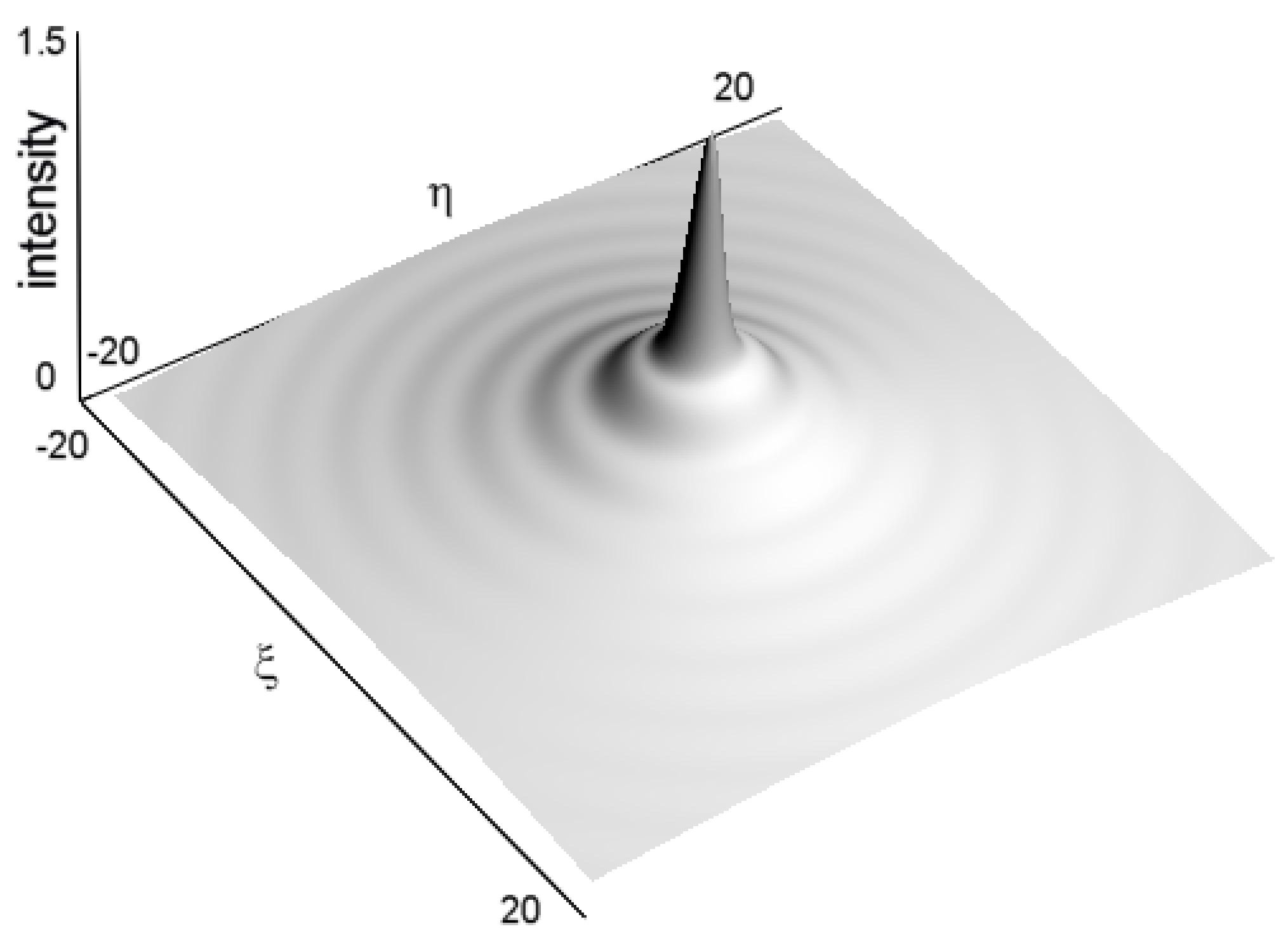}
\includegraphics[width=5cm]{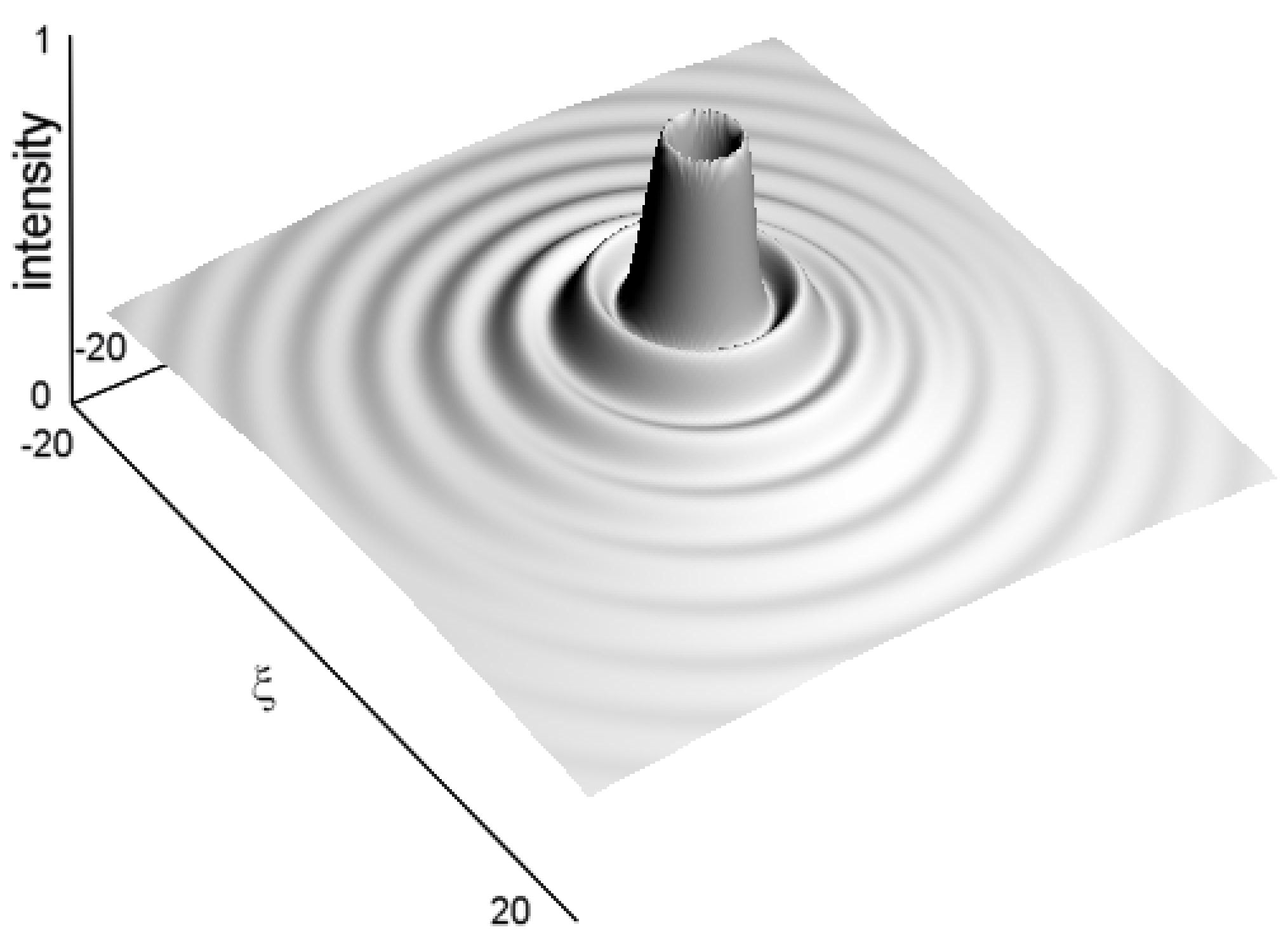}
\includegraphics[width=5cm]{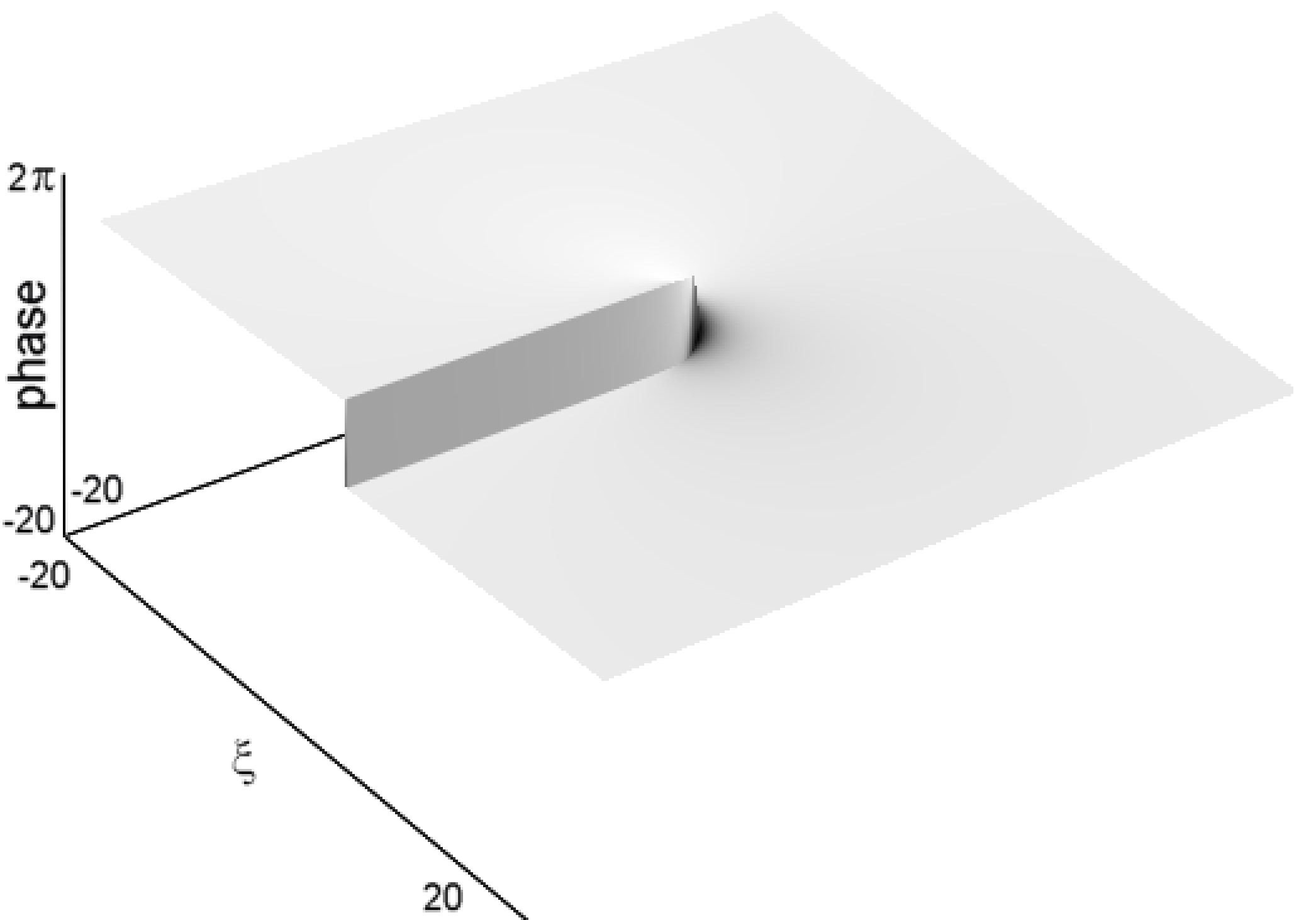}
\includegraphics[width=5cm]{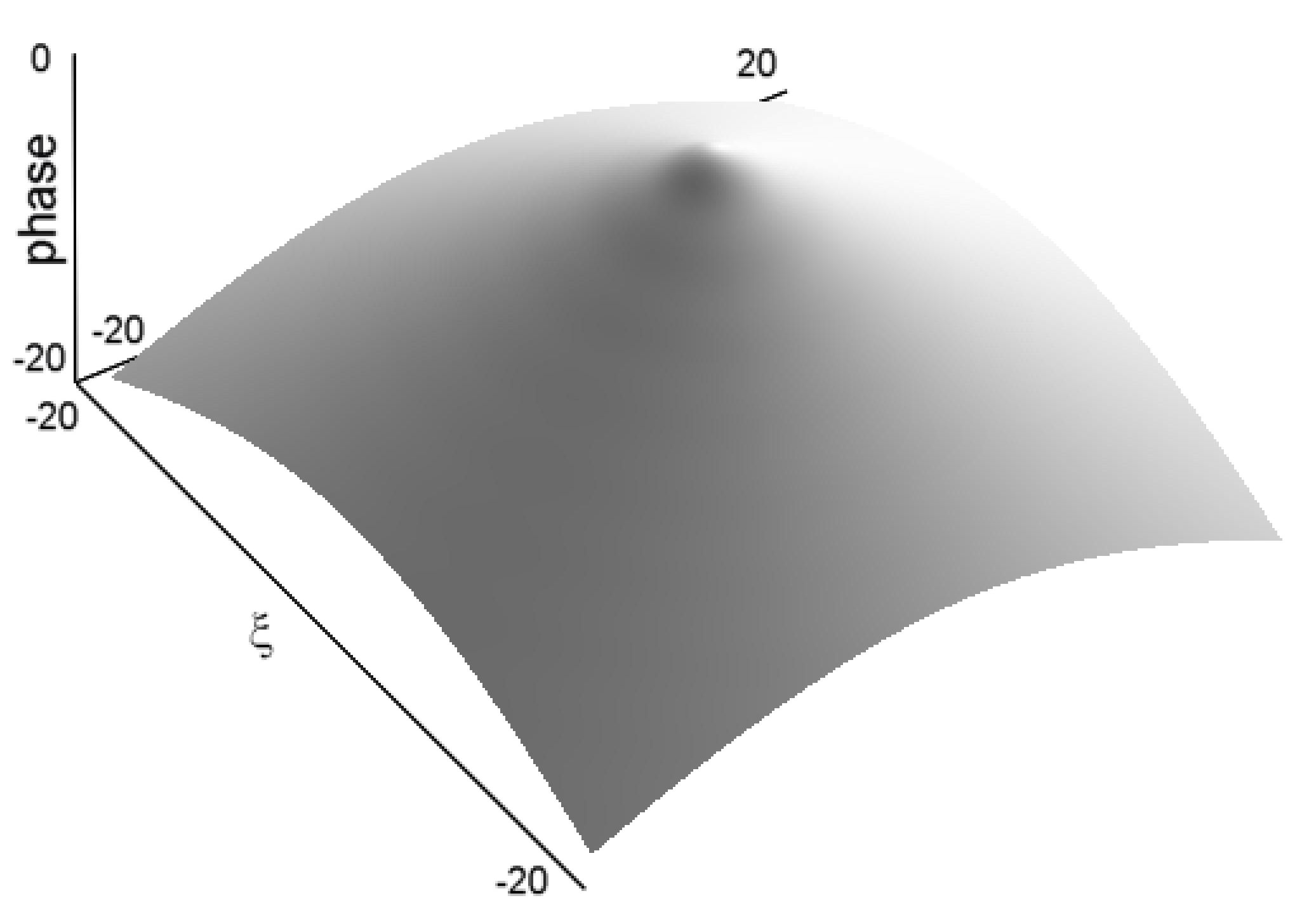}
\includegraphics[width=5cm]{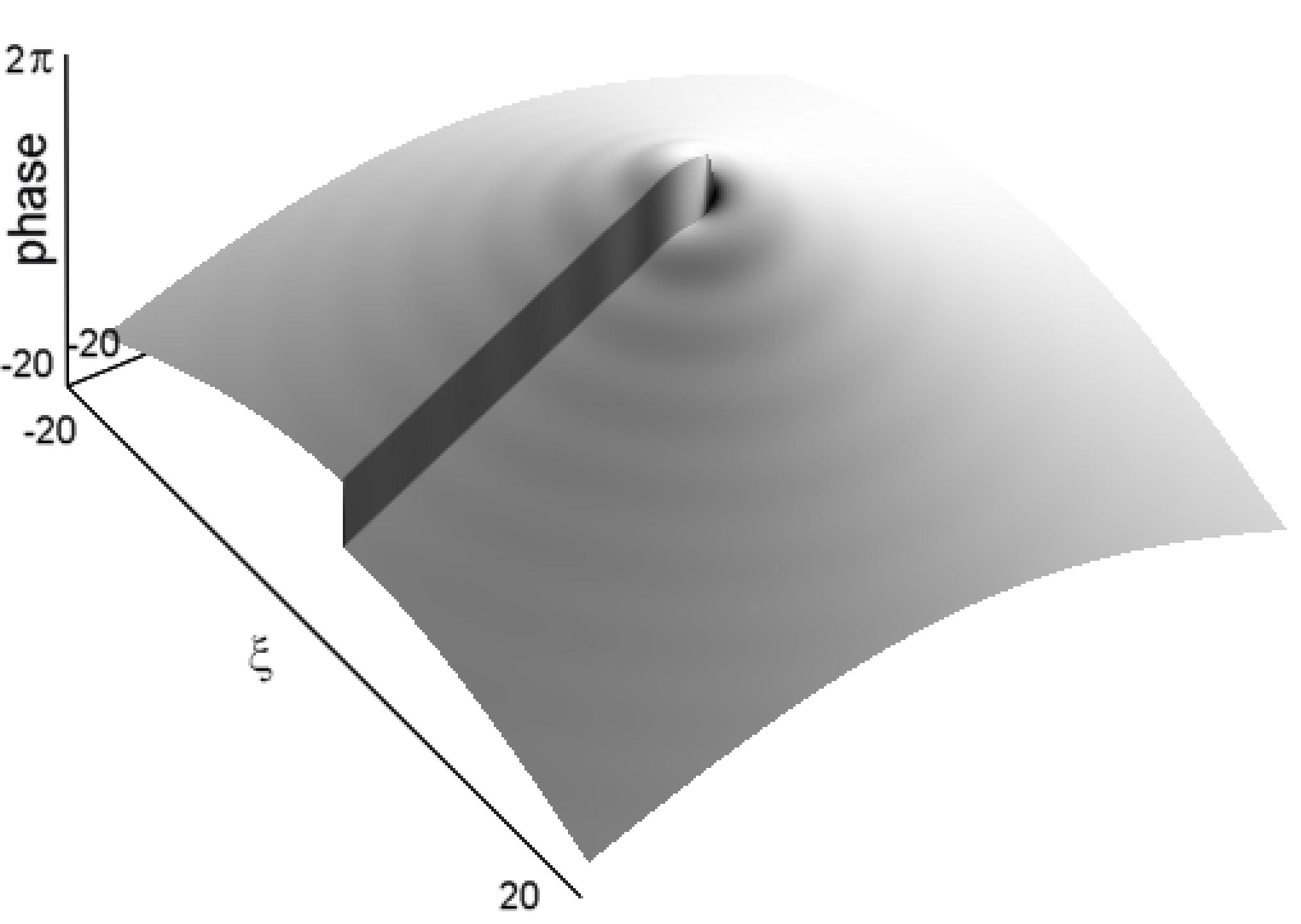}
\includegraphics[width=4.5cm]{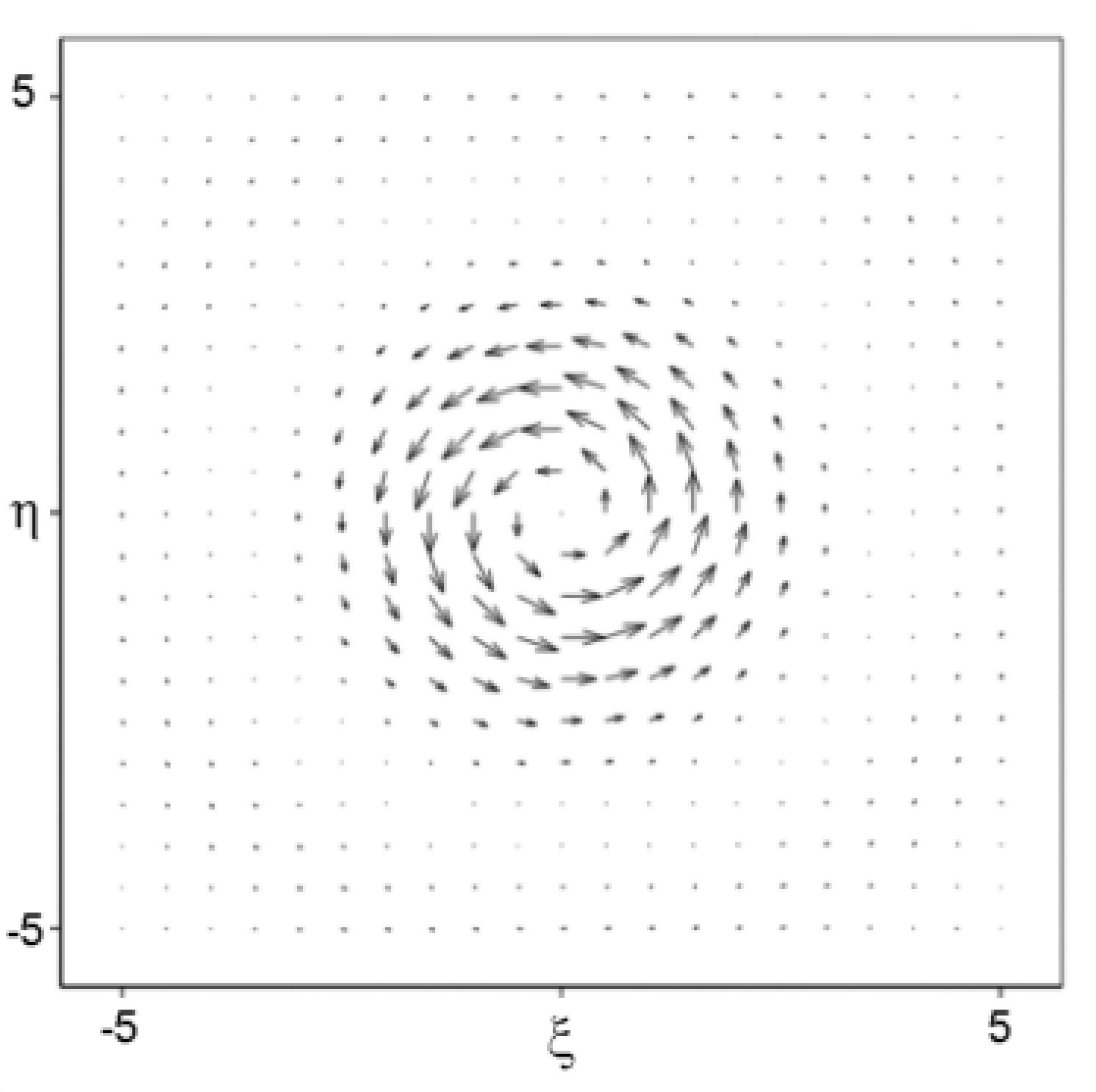}\hspace{0.5cm}
\includegraphics[width=4.5cm]{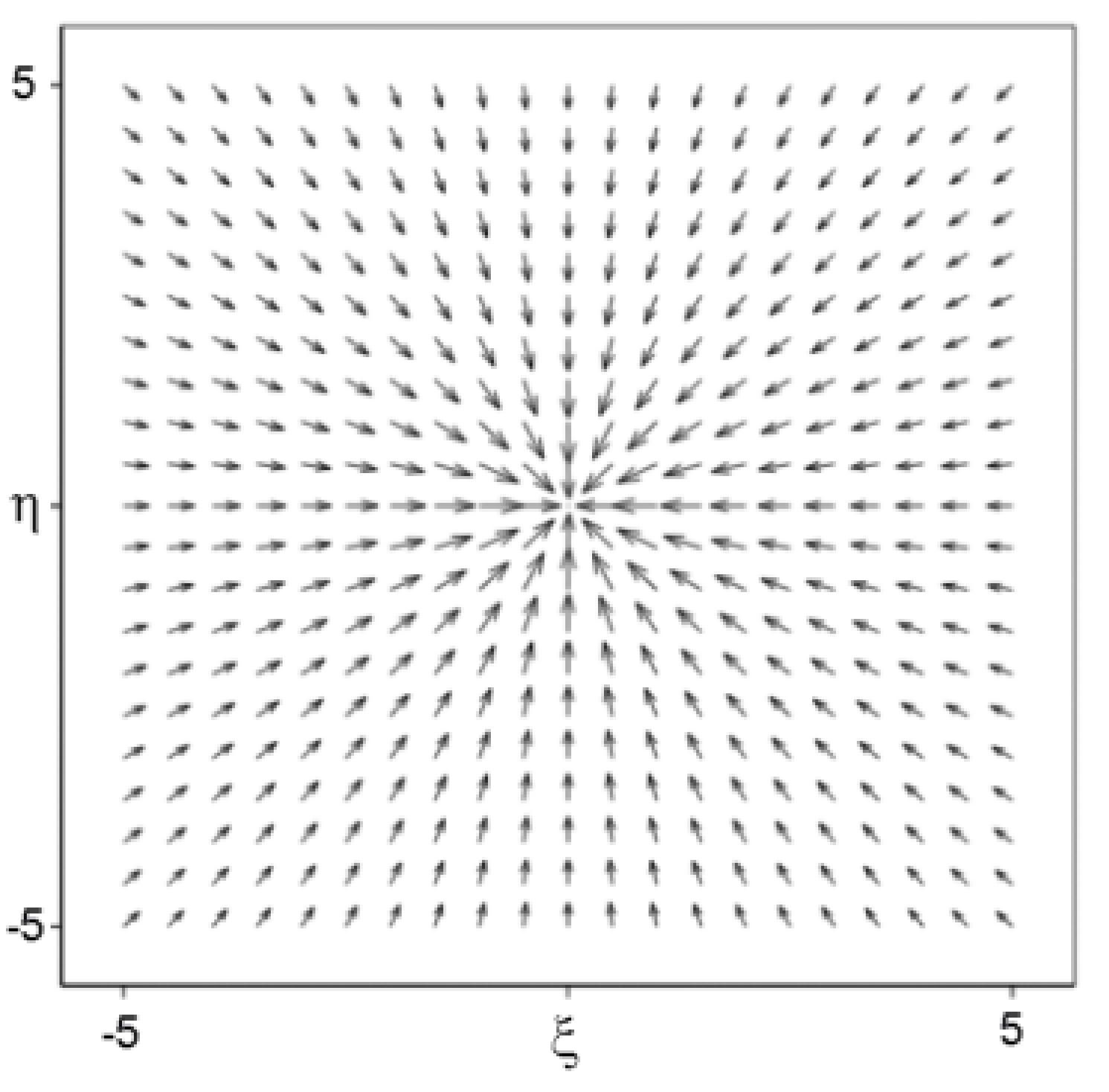}\hspace{0.5cm}
\includegraphics[width=4.5cm]{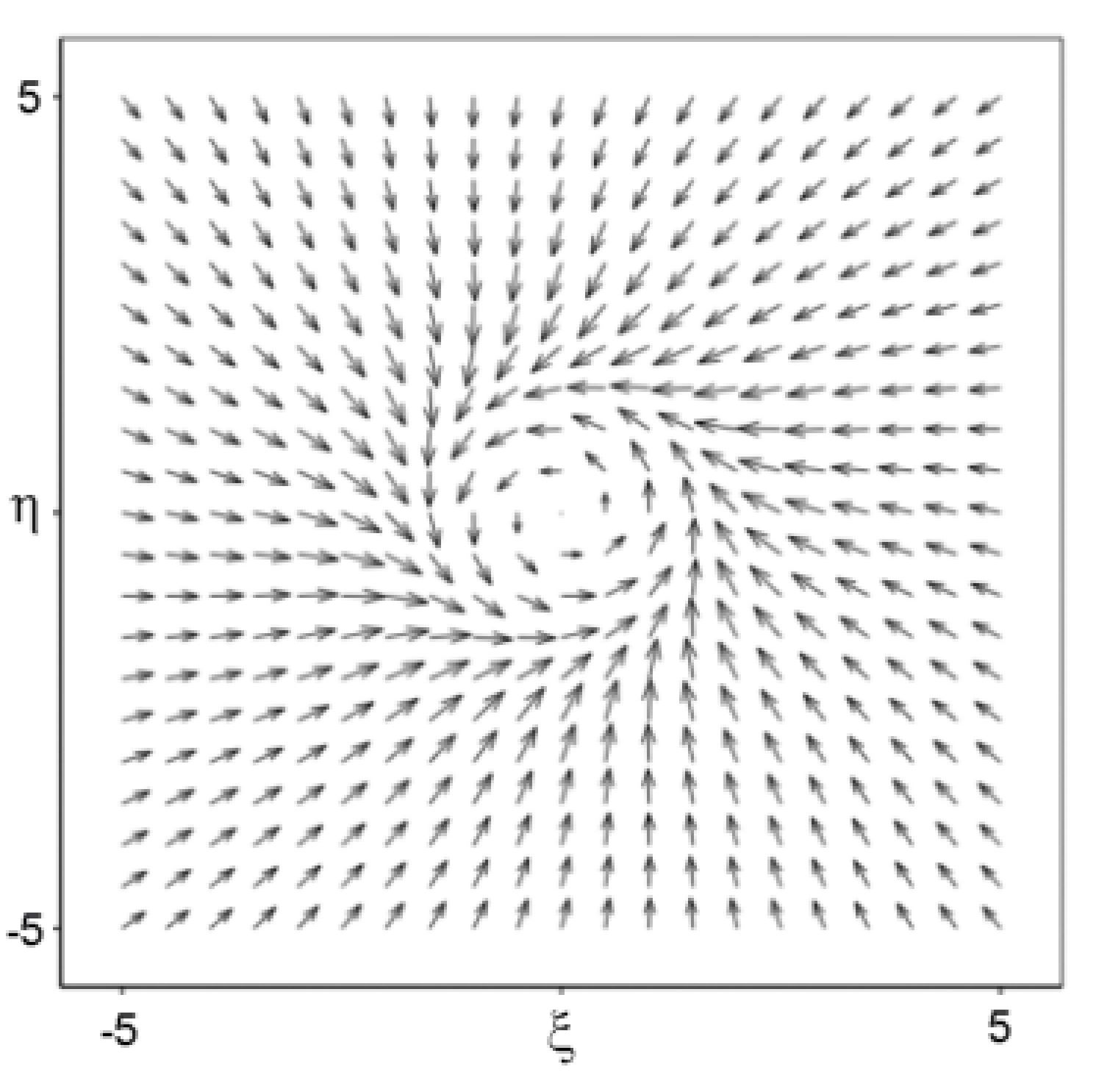}
\end{center}
\caption{\label{Fig5} Intensity profiles (first row), phase profiles (second row) and intensity current (third row) of the vortex linear BB with $l=1$, and $b_0=1.666$ (first column), of the vortex-less ($l=0$) NL-UBB with $M=4$, $\alpha_2=\alpha_4=0$ and $b_0=1.174$ (second column) and of the vortex NL-UBB with $l=1$, $M=4$, $\alpha_2=\alpha_4=0$ and $b_0=1.666$ (third column). $\xi$ and $\eta$ are normalized cartesian coordinates $\xi=\sqrt{2k|\delta|}x$ and $\eta=\sqrt{2k|\delta|}y$ in the transversal plane.}
\end{figure*}

\subsection{Stationarity with nonlinear losses of energy and orbital angular momentum}

Integration in $\rho$ in Eq. (\ref{PHAS}) leads to
\begin{equation}\label{REFILLINGE}
-2\pi\rho \frac{d\phi}{d\rho}\tilde a^2 = 2\pi \int_0^{\rho} d\rho\rho \tilde a^{2M}
\end{equation}
or $-F(\rho)=N(\rho)$ for short. This relation expresses that the nonlinear power losses in any circle of radius $\rho$, $N(\rho)$, are refuelled by an inward power flux $F(\rho)$ crossing its circumference and coming from a power reservoir at large radial distance. This is substantially the mechanism of stationarity of NL-UBB, which is only possible in beams carrying infinite power as conical beams, and applies equally to vortex NL-UBBs.

A closer look reveals some differences. Rewriting the NLSE in Eq. (\ref{NLSEN}) in terms of the amplitude and phase ($\tilde A=\tilde a e^{i\Phi}$) of a light beam (not necessarily stationary) leads to the continuity equation $(1/2)\partial_\zeta \tilde a^2 + \nabla_\perp \cdot \mathbf{j} = -\tilde a^{2M}$ for the intensity $\tilde a^2$, where the transversal current of the intensity is given by $\mathbf{j}=\tilde a^2 \nabla_\perp \Phi$ and where the nonlinear losses density $\tilde a^{2M}$ act as a distributed sink. The general condition of stationarity for the intensity is then
\begin{equation}\label{REFILLINGDIF}
-\nabla_\perp\cdot \mathbf{j}=\tilde a^{2M}\, ,
\end{equation}
expressing, by virtue of the divergence theorem, that the power losses in any finite region of the transversal plane is refuelled by an inward power flux through its contour. The above Eq. (\ref{PHAS}) is the expression of this fact for the particular case of stationarity for beams with radially symmetric intensity $\tilde a^2(\rho)$ and phase $\Phi=\phi(\rho)+l\varphi \pm \zeta$, in which case
\begin{equation}
\mathbf{j} = \tilde a^2 \left(\frac{d\phi}{d\rho}\mathbf{u}_\rho + \frac{l}{\rho}\mathbf{u}_\varphi \right)\, ,
\end{equation}
and Eq. (\ref{REFILLINGE}) is obtained by integrating Eq. (\ref{REFILLINGDIF}) over a circle of radius $\rho$. For comparison, Fig. \ref{Fig5} shows the intensity, phase and the intensity current for previously known stationary beams and for a vortex NL-UBB. For vortex BBs (and also for vortex solitons) in transparent media, the intensity current $\mathbf{j}=(\tilde a^2 l/\rho)\mathbf{u}_\varphi$ is azimuthal and solenoidal ($\nabla_\perp \cdot \mathbf{j}=0$). For the fundamental NL-UBBs, the transversal intensity current $\mathbf{j}=(\tilde a^2d\phi/d\rho)\mathbf{u}_\rho$ is radial inwards with a divergence that equals the nonlinear losses density $\tilde a^{2M}$. For vortex NL-UBBs the intensity current has a solenoidal azimuthal component and a non-solenoidal radial inwards component, so that the power spirals inwards permanently from the power reservoir at infinity towards the inner rings, where most of nonlinear power losses take place.

Unlike the fundamental UBB, the vortex NL-UBB carries an axial orbital angular momentum with a density \cite{SL} $L=\tilde a^2\partial\Phi/\partial\varphi=l\tilde a^2$ that is proportional to the intensity and to the topological charge. Thus, as the intensity, the stationary angular momentum density is permanently flowing spirally with a current
\begin{equation}
l\mathbf{j} = l\tilde a^2 \left(\frac{d\phi}{d\rho}\mathbf{u}_\rho + \frac{l}{\rho}\mathbf{u}_\varphi \right)\, .
\end{equation}
proportional to that of the intensity, and this current refills continuously the angular momentum losses $l\tilde a^{2M}$ associated to the nonlinear absorption process according to the continuity equation $-\nabla_\perp \cdot (l \mathbf{j})= l\tilde a^{2M}$ for the angular momentum density. In particular, integration over a circle of radius $\rho$ leads to
\begin{equation}
-2\pi\rho l \frac{d\phi}{d\rho}\tilde a^2 = 2\pi \int_0^{\rho} d\rho\rho l \tilde a^{2M}\, ,
\end{equation}
which expresses, by analogy with Eq. (\ref{REFILLINGE}), that the radial flux of angular momentum through any circle of radius $\rho$ refills the angular momentum losses within that circle, an angular momentum that is transferred to the material medium.

\subsection{Asymptotic behavior at large radius}

\begin{figure}
\begin{center}
\includegraphics[width=6.2cm]{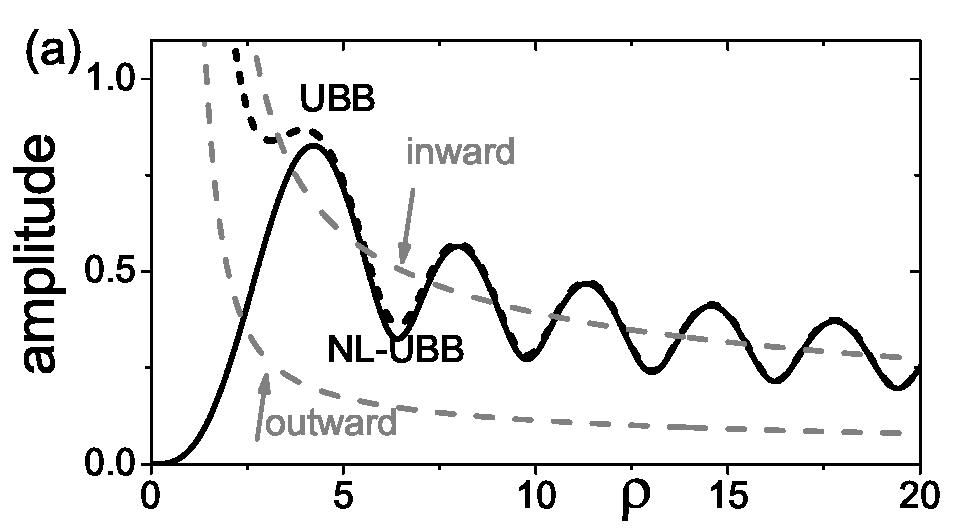}
\includegraphics[width=6.1cm]{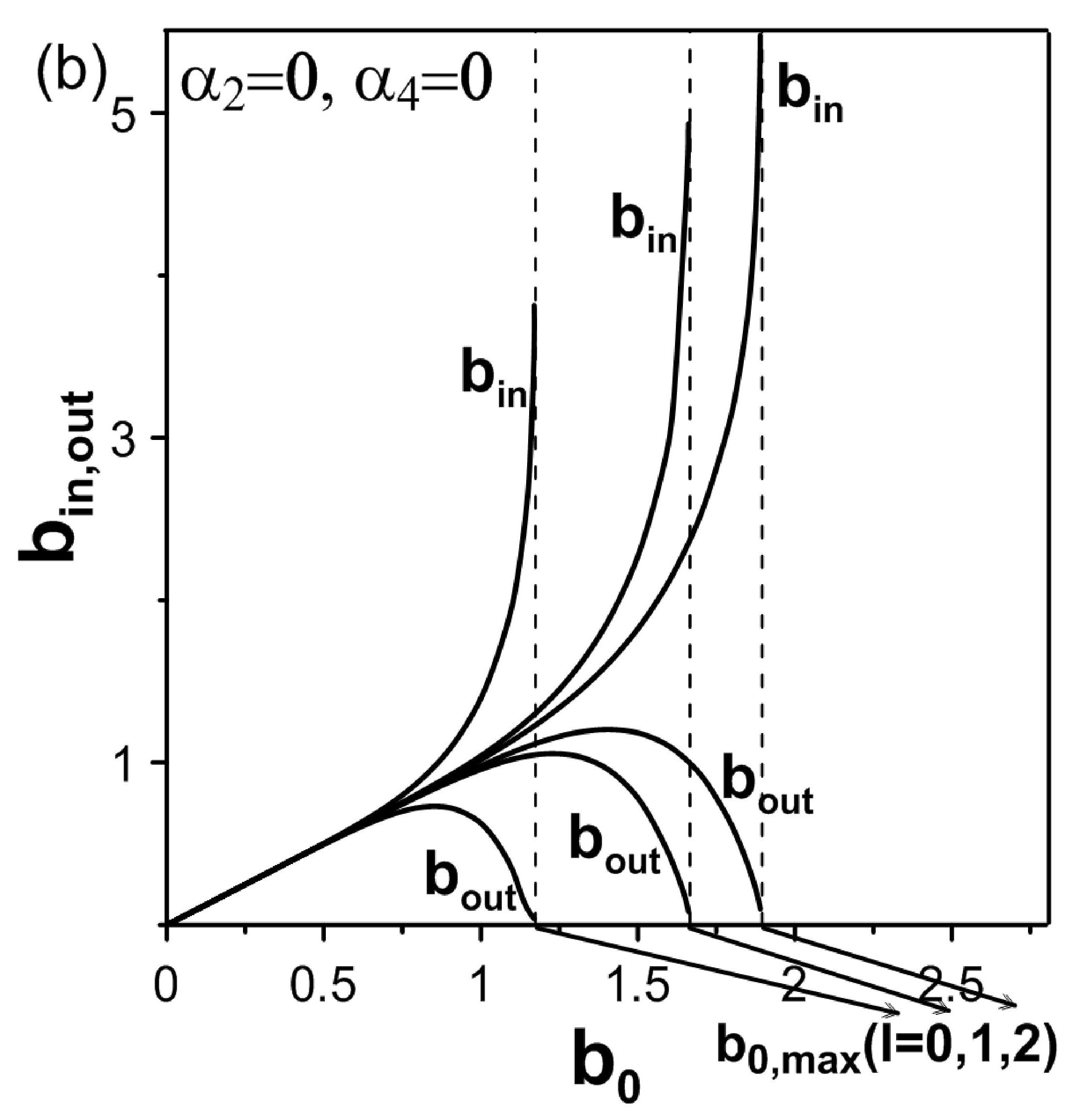}
\includegraphics[width=6.1cm]{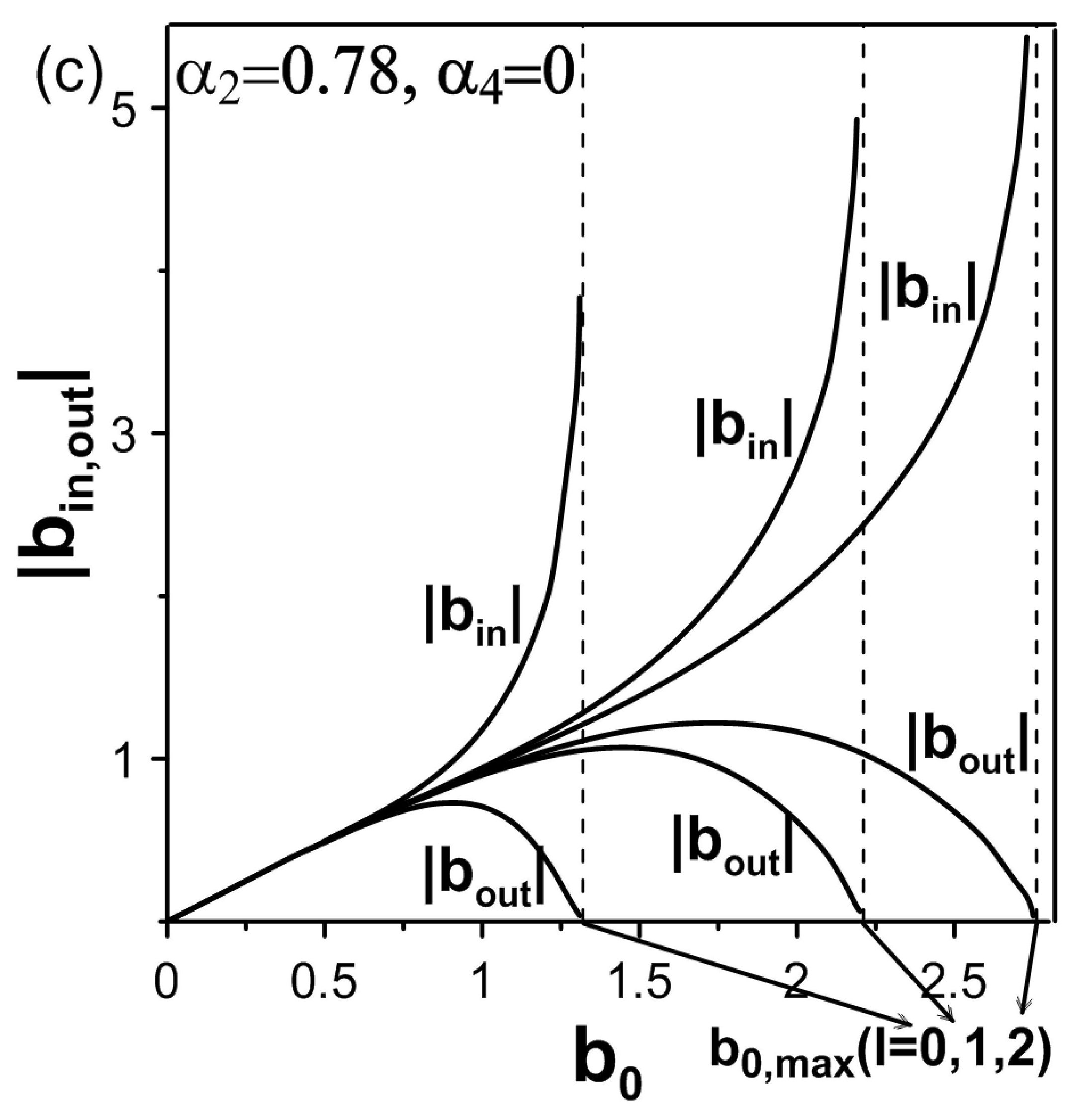}
\end{center}
\caption{\label{Fig6} (a) Amplitude of a NL-UBB ($M=4$, $\alpha_2=\alpha_4=0$, $l=3$ and $b_0=1.9$) (solid curve), its asymptotic linear UBB (dashed curve), and its H\"ankel inward and outward components (gray dashed curves). (b, c) Values of $|b_{\rm out}|$ and $|b_{\rm in}|$ for NL-UBBs in  the indicated values of $\alpha_2$, $\alpha_4$, $l$ and $M=4$, as functions of their amplitude $b_0$.}
\end{figure}

A complete understanding of the reshaping of input high-order BBs into vortex NL-UBBs requires, as we will see in Sec. \ref{SELECTION}, a detailed analysis of the structure of vortex NL-UBBs at large radial distances. Vortex NL-UBBs are linear not only at the vortex center, but also at large radial distances. At the vortex core the NL-UBB behaves as the linear Bessel beam $\tilde A\simeq b_0J_l(\rho)e^{il\varphi}e^{-i\zeta}$, or what is the same,
\begin{equation}
\tilde A\simeq \frac{1}{2}\left[b_0 H^{(1)}_l(\rho)+ b_0H^{(2)}_l(\rho)\right]e^{il\varphi}e^{-i\zeta}\, ,
\end{equation}
where the high-order BB of amplitude $b_0$ is decomposed in a ``balanced" superposition of two high-order H\"ankel beams of equal amplitudes $b_0$, the H\"ankel beam $H^{(1)}_l(\rho)e^{il\varphi}e^{-i\zeta}$ carrying power spirally outwards, and the H\"ankel beam $H^{(2)}_l(\rho)e^{il\varphi}e^{-i\zeta}$ carrying the same amount of power spirally outwards \cite{SALO}. As the fundamental NL-UBB \cite{PORRAS1}, a vortex NL-UBB behaves at large radius as the unbalanced high-order Bessel beam (UBB)
\begin{equation}\label{ASYMP}
\tilde A \simeq \frac{1}{2}\left[b_{\rm out} H^{(1)}_l(\rho)+ b_{\rm in}H^{(2)}_l(\rho)\right]e^{il\varphi}e^{-i\zeta}\, ,
\end{equation}
[Fig. \ref{Fig6}(a),solid and dashed curves] where the two interfering high-order H\"ankel beams have different amplitudes $b_{\rm out}$ and $b_{\rm in}$ [Fig. \ref{Fig6}(a),gray dashed curves]. Using that $H^{(1,2)}_l(\rho)\simeq\sqrt{2/(\pi z)}\, e^{\pm i[\rho-(\pi/2)(l-1/2)]}$ at large $\rho$ \cite{GRADSHTEYN}, the condition that the inward radial flux $-F(\rho)=-2\pi\rho \tilde a^2 d\phi/d\rho=2\pi\rho \mbox{Im}[\tilde A(\partial \tilde A^\star/\partial\rho)]$ matches at large $\rho$ the total nonlinear losses $N(\infty)$ is readily seen to impose the constraint
\begin{equation}\label{REFILLING2}
|b_{\rm in}|^2-|b_{\rm out}|^2=N(\infty)\, ,
\end{equation}
between the amplitudes $b_{\rm in}$ and $b_{\rm out}$, from which it follows that $|b_{\rm in}|>|b_{\rm out}|$ for any vortex NL-UBB. Also, the radial intensity at large $\rho$ is given by
\begin{equation}\label{MEAN}
\tilde a^2 =\frac{1}{2\pi\rho}\left\{|b_{\rm out}|^2\!+\!|b_{\rm in}|^2 \!+\!
            2|b_{\rm out}b_{\rm in}|\cos\!\left[2\rho+\kappa \right]\right\}
\end{equation}
with $\kappa=-\pi l-\pi/2+\mbox{arg}(b_{\rm out}/b_{\rm in})$. Thus, $2\pi\rho \tilde a^2$ represents harmonic oscillations of contrast $C=2|b_{\rm in}b_{\rm out}|/(|b_{\rm in}|^2+|b_{\rm out}|^2)$ about the mean value $|b_{\rm in}|^2+|b_{\rm out}|^2$. The values of $|b_{\rm in}|$ and $|b_{\rm out}|$ for a given vortex NL-UBB with $b_0$ can be easily extracted from the above properties and its numerically evaluated profile, and are represented as functions of $b_0$ in Fig. \ref{Fig6}(b) and (c) for NL-UBBs with different charges in different media. It turns out that $b_{\rm in}$ and $b_{\rm out}$ are real in absence of dispersive nonlinearities, and $\mbox{arg} b_{\rm in}=-\mbox{arg} b_{\rm out}$ with dispersive nonlinearities as self-focusing or self-defocusing.

\section{On the ``selection problem" in the dynamics of high-order Bessel beams} \label{SELECTION}

In Sec. \ref{DYNAMICS1} we showed that a high-order linear BB survives as a stationary beam in a nonlinear medium with nonlinear absorption by reaching spontaneously a new propagation invariant state in the form of a vortex NL-UBB of same cone angle and topological charge as the input BB, but in general of much lower intensity. With the notation introduced in Sec. \ref{STATIONARY}, the input BB specified by a certain value of $b_0$, say $b_0(0)$, transforms at large $\zeta$ into a NL-UBB specified by a different value of $b_0$, say $b_0(\infty)$ [e. g., the BB with $b_0(0)=3$ transforms into the NL-UBB with $b_0(\infty)=1.60$ in the simulation of Fig. \ref{Fig1}]. For a complete description of the dynamics, the question that remains unsolved is to determine which particular NL-UBB is selected as the final stationary state given the input BB, i. e., the value of $b_0(\infty)$ given the value of $b_0(0)$, and to understand why. This problem arose in Ref. \cite{CONE} and remained unsolved in relation of vortex-less BBs and NL-UBBs, where the peak intensities of all input BBs and final NL-UBBs are normalized to unity, but their ratio is actually $b_0^2(0)/b_0^2(\infty)$.

\begin{figure}[b]
\begin{center}
\includegraphics[width=4.2cm]{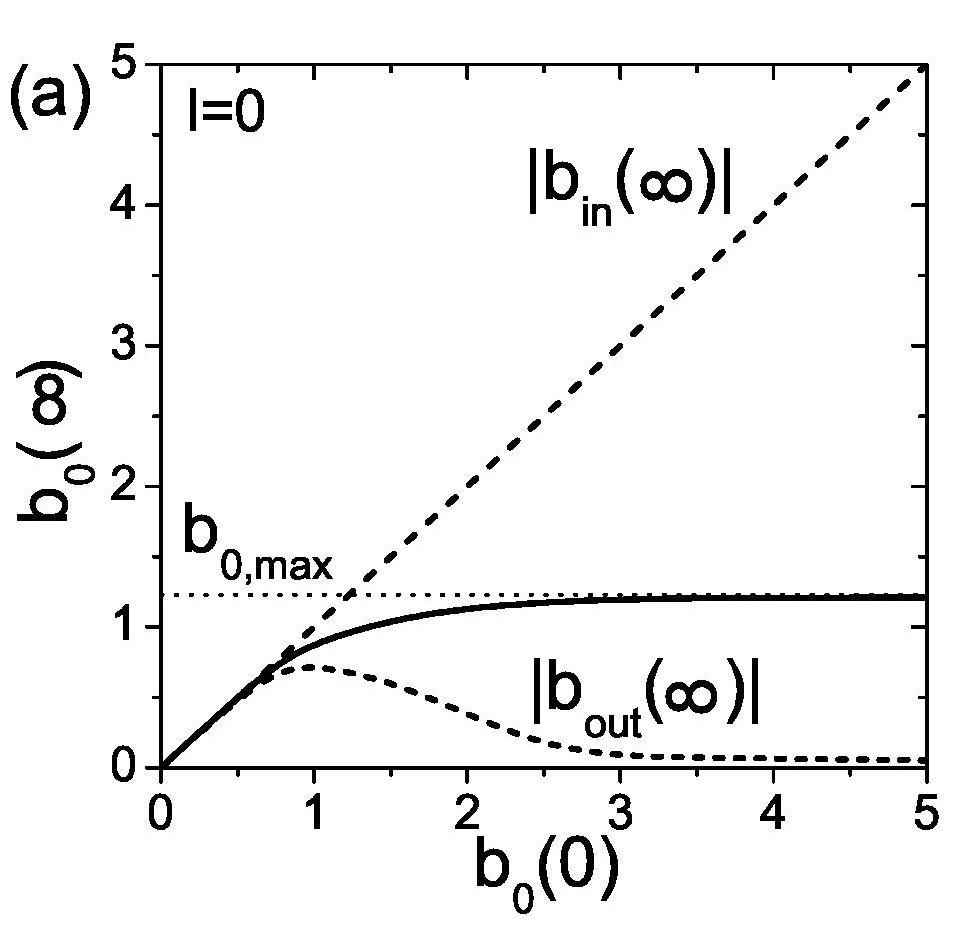}
\includegraphics[width=4.2cm]{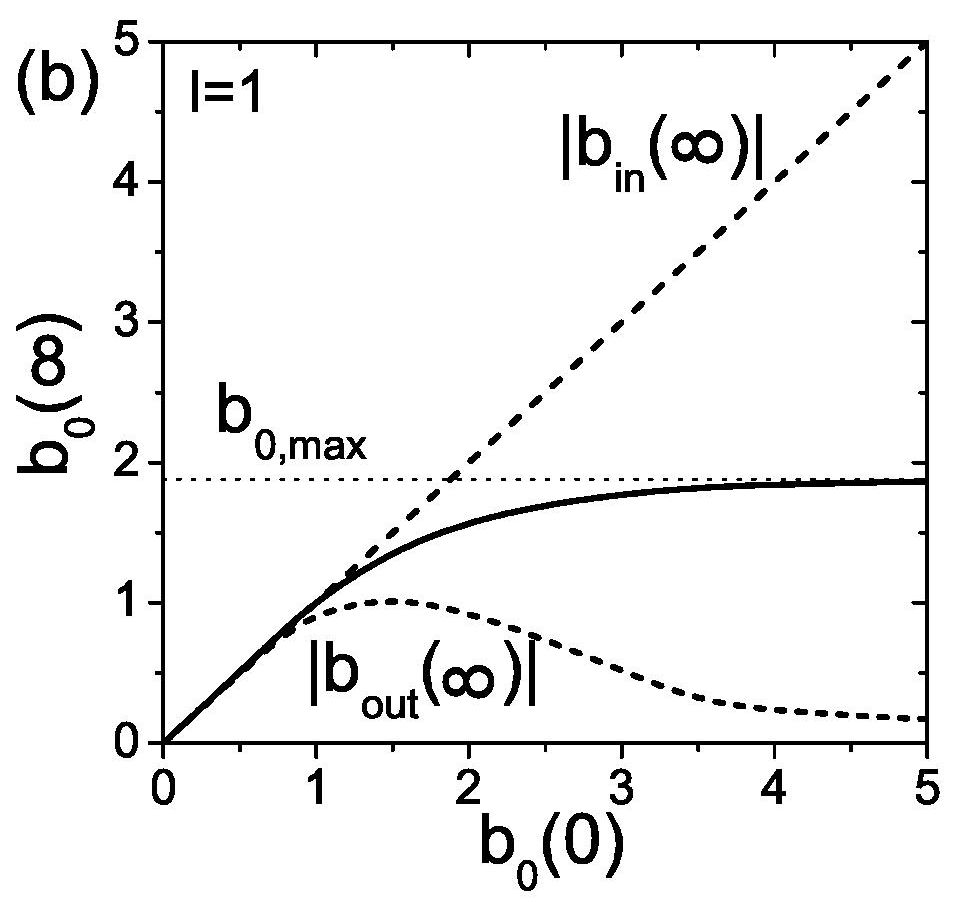}
\end{center}
\caption{\label{Fig7} Values of $b_0(\infty)$ of the final vortex NL-UBB as a function of $b_0(0)$ of the input vortex BB for $l=0$ and $l=1$ in a medium with $M=4$, $\alpha_2=0.5$ and $\alpha_2=-0.25$, obtained by solving numerically the NLSE in Eq. (\ref{NLSEN}). The dashed curves represent the amplitudes $|b_{\rm in}(\infty)|$ and $|b_{\rm out}(\infty)|$ of the inward and outward H\"ankel components of the final vortex NL-UBB with the amplitude $b_0(\infty)$. As a function of $b_0(0)$, $|b_{\rm in}(\infty)|$ is found numerically to be the identity function, i. e., $|b_{\rm in}(\infty)|=b_0(0)$.}
\end{figure}

In a purely numerical approach, the values of $b_0(\infty)$ for each $b_0(0)$ can be found by solving numerically the NLSE in Eq. (\ref{NLSEN}), as done in Sec. \ref{DYNAMICS1}. The pairs [$b_0(0),b_0(\infty)]$ extracted from these simulations are displayed in Fig. \ref{Fig7} for $l=0,1$ in a particular medium. A first conclusion is that the final NL-UBB approaches that of maximum amplitude supported by the medium at very high intensities of the input BB, i. e., $b_0(\infty)\rightarrow b_{0,\rm max}$ when $b_0(0)\gg b_{0,\rm max}$. In the opposite side, the selected NL-UBB does not differ substantially from the launched BB when the intensity is low enough so that NLL are negligible, i. e., $b_0(\infty)\rightarrow b_0(0)$ when $b_0(0)\rightarrow 0$. At intermediate amplitudes, a certain NL-UBB with $b_0(\infty)$ between $0$ and $b_{0,\rm max}$ is selected.

The law of the NL-UBB selection is suggested by the dashed lines in Fig. \ref{Fig7}. They represent the amplitudes $|b_{\rm in}(\infty)|$ and $|b_{\rm out}(\infty)|$ of the asymptotic inward and outward H\"ankel components of the final NL-UBB. It turns out from these numerical simulations that $|b_{\rm in}(\infty)|=b_0(0)$. This result is seen to hold irrespective of the medium (the values of $M$ and the dispersive nonlinearities) and the topological charge $l$. Since for the input BB $|b_{\rm in}(0)|=b_0(0)$ too, we infer the law that {\em the amplitude of the asymptotic inward H\"ankel component is conserved in the nonlinear propagation,} and only the amplitude of the outward H\"ankel component diminishes. The inward H\"ankel beam is indeed a linear beam that continuously brings power from the energy and angular momentum reservoir at infinity, and therefore is not affected by the nonlinear effects that take place in the inner rings. Thus, given the value of $b_0(0)$ of the input BB, the NL-UBB selected as the final stage of the dynamics is that with $b_0(\infty)$ whose inward component equals $b_0(0)$, that is,
\begin{equation}
|b_{\rm in}(\infty)|=b_0(0)\, .
\end{equation}
Figures such as Figs.\ref{Fig6}(b) and (c) for the stationary NL-UBBs provide therefore a solution to the selection problem. Given a medium ($M$ and the dispersive nonlinearities) and charge $l$, $b_0(\infty)$ is the value of the abscissa that intersects the curve $|b_{\rm in}|$ at the ordinate value $b_0(0)$.
\begin{figure}[t]
\begin{center}
\includegraphics[width=7cm]{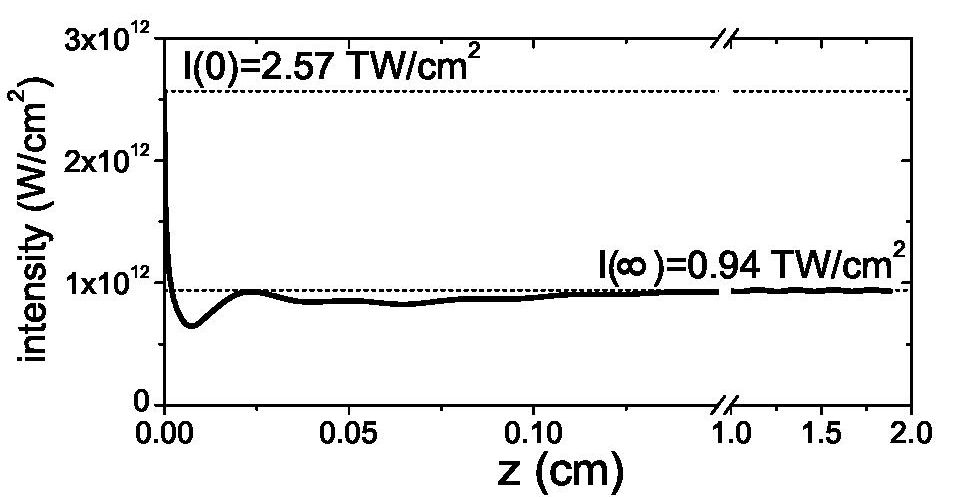}
\end{center}
\caption{\label{Fig8} Peak intensity of a zero-order BB launched in fused silica as a function of propagation distance, obtained by numerical solution of Eq. (\ref{NLSE}). Details of the input BB and of the optical properties of fused silica are given in the text.}
\end{figure}

As an example, and to express these results with physical magnitudes, we consider the input zero-order BB $A(r,0)=\sqrt{I(0)}J_0(k\theta r)$ at $\lambda=527$ nm carrier wave length ($\omega=3.58$ fs$^{-1}$) and of cone angle $\theta=1^\circ$ in fused silica, considered as a pure cubic Kerr medium with four-photon absorption ($n=1.461$, $k=1.742\times 10^5$ cm$^{-1}$, $n_2=2.7\times 10^{-16}$ cm$^2$/W, $M=4$, and $\beta^{(4)}=2\times 10^{-34}$ cm$^5$/W$^3$ \cite{SF}, so that $\alpha_2=0.78$, $\alpha_4=0$). For the input intensity $I(0)=2.57$ TW/cm$^2$, the dimensionless amplitude is $b_0(0)\simeq 2.0$. The final stationary state is then the NL-UBB with dimensionless amplitude $b_0(\infty)$ such that its inward H\"ankel component is $|b_{\rm in}(\infty)|=2.0$. Unfortunately, there is no simple (analytical) relation between $|b_{\rm in}|$ and $b_0$ for NL-UBBs, and a numerical analysis of the stationary states for the involved values of $M$, $l$, and $\alpha_j$ is necessary. This analysis is performed in Fig.  \ref{Fig6}(c), showing that the NL-UBB with $|b_{\rm in}(\infty)|=2.0$ is that with $b_0(\infty)=1.21$, corresponding to a peak intensity $I(\infty)= 0.94\times 10^{12}$ W/cm$^2$. As a verification, Fig. \ref{Fig8} depicts the peak intensity of the input BB as it propagates in fused silica as obtained from numerical solution of the NLSE in Eq. (\ref{NLSE}). After a sudden diminution, the peak intensity is seen to stabilize into the value predicted from the above theory.

\section{Conclusions}

We have reported on the existence of non-diffracting and non-attenuating vortex beams supported by nonlinear absorption. These vortex NL-UBB form narrow and (ideally) infinitely long tubes of light where energy and orbital angular momentum can be transferred to matter, but the beam energy and angular momentum are continuously restored by spiral currents coming from a reservoir at large radial distances. In real settings, a vortex NL-UBB can be formed only up to a finite radial distance, say $r_{\rm max}$, by which the reservoir is of limited capacity. It will then be depleted, and the refilling mechanism will cease, after a distance of the order of the diffraction-free distance $z_{\rm free}=r_{\rm max}/\theta$, as described for other conical light beams \cite{BB}. Vortex NL-UBBs arise naturally in the propagation of linear BBs at intensities at which multiphoton absorption and possibly other nonlinearities are significant. In this sense, we have provided an accurate description of the nonlinear propagation of (vortex or vortex-less) Bessel beams in nonlinear media in terms of their transformation into the NL-UBB that preserve the cone angle, the topological charge and the inward current coming from the reservoir.

\section{Acknowledgments}

M. A. P. acknowledges financial support from Project MTM2012-39101-C02-01 of Ministerio de Econom\'{\i}a y Competitividad of Spain.

\end{document}